\documentclass{article}

\usepackage{arxiv}

\usepackage[utf8]{inputenc} 
\usepackage[T1]{fontenc}    
\usepackage{hyperref}       
\usepackage{url}            
\usepackage{booktabs}       
\usepackage{amsfonts}       
\usepackage{nicefrac}       
\usepackage{microtype}      
\usepackage{lipsum}
\usepackage{natbib}
\usepackage{amsmath}
\usepackage{graphicx}

\newcommand{\tbeta}{\boldsymbol{\beta}}
\newcommand{\tx}{\boldsymbol{x}}

\title{How are emergent constraints quantifying uncertainty and what do they leave behind?}

\author{
Daniel B. Williamson\thanks{Also at The Alan Turing Institute for Data Science and Artificial Intelligence}\\
Department of Mathematical Sciences, \\
University of Exeter, UK \\
\texttt{d.williamson@exeter.ac.uk}
\And Philip G. Sansom \\
Department of Mathematical Sciences, 
\\University of Exeter, UK \\
\texttt{p.g.sansom@exeter.ac.uk}
}

\begin{document}
\maketitle

\begin{abstract}
The use of emergent constraints to quantify uncertainty for key policy relevant quantities such as Equilibrium Climate Sensitivity (ECS) has become increasingly widespread in recent years. Many researchers, however, claim that emergent constraints are inappropriate or even under-report uncertainty. In this paper we contribute to this discussion by examining the emergent constraints methodology in terms of its underpinning statistical assumptions. We argue that the existing frameworks are based on indefensible assumptions, then show how weakening them leads to a more transparent Bayesian framework wherein hitherto ignored sources of uncertainty, such as how reality might differ from models, can be quantified. We present a guided framework for the quantification of additional uncertainties that is linked to the confidence we can have in the underpinning physical arguments for using linear constraints. We provide a software tool for implementing our general framework for emergent constraints and use it to illustrate the framework on a number of recent emergent constraints for ECS. We find that the robustness of any constraint to additional uncertainties depends strongly on the confidence we can have in the underpinning physics, allowing a future framing of the debate over the validity of a particular constraint around the underlying physical arguments, rather than statistical assumptions.
\end{abstract}


\section{Introduction}
Emergent constraints have become a popular and controversial topic within the climate science community over the last number of years \citep{hallqu06,wenzeletal16,coxetal18}. For some policy relevant quantity that we cannot observe now, for example Equilibrium Climate Sensitivity (ECS), researchers seek to discover whether there are observations that we can make that would quantify or constrain our uncertainty in that quantity.

To answer this question, the community has looked to the ensembles of the coupled model intercomparison projects CMIP3 \citep{meehletal07} and CMIP5 \citep{tayloretal12}, and now CMIP6 \citep{eyringetal16}. The idea is to find a (typically linear) ``emergent'' relationship across the models between the quantity of interest (QoI, for example ECS) and something that can be measured. For example, \citet{hallqu06} found that the current seasonal cycle had a linear relationship with snow albedo feedback in CMIP models. \citet{coxetal18} relates ECS to a particular metric of climate variability. Once such a relationship is found, the models are used to estimate it via regression. Observations from the real world, coupled with the regression produce a constraint on the QoI in reality.

There are a number of reasons that this practice has caused controversy. One is the way in which the constraints are found. Some use physical reasoning to show we would expect a linear relationship between model quantities, and then look to confirm this through the ensemble \citep[e.g.][]{coxetal18}. Others have suggested data mining be used to find them \citep[e.g.][]{karpechkoetal13}. We discuss these ideas later. Another source of controversy is the simplicity of the treatment versus the complexity of the models and the quantities of interest. The argument is that the observed relationships are not ``emergent'' from the physics and hence predictive, but a result of the interaction of many different processes, well captured in the models or not, which must be better understood in order to say something about reality. A final concern is that emergent constraints actually under-estimate uncertainty. \citet{bowmanetal18} show that older methods have indeed under-represented uncertainty by constructing a statistical framework for emergent constraints that properly accounts for the uncertainty in the observations themselves, though neglects other sources that we seek to address here.

In this paper we will explain the underpinning statistical assumptions and judgements that lead to the emergent constraints model.  We will highlight the different sources of uncertainty that should be present when finding emergent constraints and show where they can enter the usual framework. We will argue for a simple generalisation to existing methods that allows hitherto neglected uncertainties to be quantified and then compare results from this extended model to existing results in the literature. Our goal is to translate the existing underpinning statistical assumptions behind emergent constraints and then place them in a more general framework that allows all assumptions for any emergent constraint analysis to be transparently understood. Our framework highlights all sources of uncertainty and offers methodology for guided quantification of these additional uncertainty sources. To accompany the paper we present a version controlled software tool capable of fitting the general emergent constraints model to user-inputted data that allows users to explore the effects of all sources of uncertainty on the analysis. Whether the statistical assumptions themselves are valid for any particular emergent constraint, or at all when using CMIP and observations in this way is a question for the climate community to resolve. This paper and its accompanying software can help to frame this discussion.

In Section~\ref{Sect:theory} we present the strong statistical assumptions behind emergent constraints and generalise the framework by weakening them. We show where key uncertainties were being ignored and show how they can be quantified going forward. In Section~\ref{sect:example1} we apply the generalised framework to the emergent constraint on ECS recently presented by \citet{coxetal18} to demonstrate the effect of acknowledging additional sources of uncertainty. In Section~\ref{sect:priors} we discuss quantifying these additional sources of uncertainty and present a default guided specification which is available to use through our software tool. In Section~\ref{sect:examples} we apply the new framework to a collection of constraints on ECS from the literature and discuss the interpretation of different emergent constraints analyses for the same quantity. Section~\ref{sect:discussion}
 contains discussion. Appendix~\ref{appendix:proofs} contains some of the mathematical results used to derive our more general framework. The software tool and user instructions are available at \url{https://github.com/ps344/emergent-constraints-shiny}.

\section{Exchangeability and emergent constraints}
\label{Sect:theory}
 
Emergent constraints are formed through relationships between climate models. Suppose we have an ensemble of climate models of size $n$. From each model we can obtain the value of a predictor (something we can observe) $x_i$, and a response (e.g. ECS), $y_i$, for $i = 1,\ldots,n$. The general concept is to use this ensemble data to fit a regression model:
\begin{equation*}
    y_i = \beta_0 + \beta_1 x_i, \quad\quad i = 1,\ldots,n,  
\end{equation*} 
and use this model to ``constrain'' uncertainty for the response in the real world, $y^*$, given a value for the predictor from the real world, $x^*$. But what kinds of assumptions are required to underpin such an approach and in what contexts might they be valid?

\subsection{Ordinary least squares and classical regression}

Least squares estimates of $\tbeta = (\beta_0,\beta_1)^T$ can be obtained without any assumptions, simply by minimising the sum of squared distances between the $y_i$ and the $\tbeta^T \tx_i$ (where $\tx_i = (1,x_i)^T$), leading to well known formulae for estimates $\hat{\tbeta}$ \citep[see, e.g.][]{drapersmithARA}. Using such estimates to account for uncertainty in other models or reality, however, requires a statistical model to formalise the underpinning assumptions.

A classical regression assumes that for the true $\tbeta$, the errors from the fit are independent and normally distributed with common variance
\begin{equation*}
  y_i = \tbeta^T \tx_i + e_i, \quad\quad e_i \sim \mathrm{N} (0, \sigma^2).
\end{equation*} 
The maximum likelihood estimator $\hat{\tbeta}$ then coincides with the least squares estimator, and prediction intervals can be constructed for unobserved models or even reality $y^*$ (at $x^*$), if they are assumed to be independent draws from the same error distribution.

The principal difficulty with this framework as it pertains to climate models is its characterisation of the uncertainties that are being quantified and its lack of flexibility. For this model, uncertainty arises through the random nature of the climate model as it deviates from the line $\tbeta^T \tx$. As the models are deterministic, this randomness can only come from initial condition uncertainty leading us to view the deviation as the result of observing a random point on each model's attractor, and the line representing the mean of the attractor as it changes with $x$. Note this implies every model's attractor has the same `variability' ($\sigma^2$), a fact we don't believe many would be minded to defend strongly. Further, the treatment of the real world as ``just another model'' following the same error distribution is unnatural and impossible to square with the classical interpretation (e.g. what is the meaning of a frequency distribution for ECS?).

More natural is a Bayesian approach in which we acknowledge that, before we observe the models, we are uncertain as to what their $x_i$ and $y_i$ values will be, just as we are uncertain about the corresponding $x^*$ and $y^*$ values for reality. The key concept here is the prior judgement of \textit{exchangeability} between the responses given the predictors. Exchangeability is a weak assumption that amounts to indifference over labels \citep{definetti1974,definetti1975}. Here it says that, for any $i$, $j$, we think that no information about the pairs ($y_i,x_i$) and $(y_j,x_j)$ is encoded in their labels $i$ and $j$. Hence if $x_i$ and $x_j$ took the same value, our distribution for $y_i$ and $y_j$ would be the same \textit{a priori}. Here the $i$ and $j$ are the labels for the different climate models so applying this assumption for an emergent constraint means that if the value of the predictor turned out to be the same for any subset of models, there is nothing else that we know about those models that would lead us to change our distribution for the response before seeing the model responses. 

Coupled with the assumption that there is a linear relationship between $x_i$ and $y_i$, this type of exchangeability implies
\begin{equation*}
  \mathrm{E} \left[ y_i \mid x_i \right]  = \tbeta^T \tx.
\end{equation*} 
Assuming that the $y_i \mid x_i$ are independent and identically distributed, as in the classical setting, trivially implies exchangeability. To make use of the weaker exchangeability assumption without assuming independence, de Finetti's representation theorem and its various generalisations \citep{hewittsavage55, diaconisfreedman80} imply that given exchangeability, there exists a probability model $p(y \mid x,\theta)$, considered to be a limit of a function of the $y_i$ and a prior distribution on $\theta$, $\pi(\theta)$ so that
\begin{equation*}
  p(y_1,\ldots,y_n \mid x_1,\ldots,x_n) = 
    \int \prod_{i=1}^n p(y_i \mid x_i,\theta) \pi(\theta) d\theta.
\end{equation*} 
We include this result for interest only. From a practical perspective it means that for exchangeable quantities we can behave as if a finite collection is an independent random sample from some probability model, parameterised with $\theta$ and with a prior distribution on $\theta$, $\pi(\theta)$.

In the case of emergent constraints, we might view the symmetry and ubiquity of the Normal distribution as attractive for our choice of distribution and set 
\begin{equation}
  \label{models}
  y_i \mid x_i, \theta \sim \mathrm{N} (\tbeta^T \tx, \sigma^2),
\end{equation} 
where $\theta = \{\tbeta, \sigma^2\}$ and we choose a prior $\pi(\tbeta,\sigma^2)$ to encode any prior information that we have. This is the Bayesian version of the regression problem and, though perhaps unusual at first, is, as argued above, based on much weaker assumptions than the classical version. What's more, if the usual so-called ``reference'' prior, $\pi(\tbeta,\sigma^2) \propto 1/\sigma^2$ is used, the classical analysis and the Bayesian analysis coincide \citep[see, e.g.][]{bernardosmith94, gelmanbda}. So we can view the current approach used to model emergent constraints as Bayesian with the reference prior on the regression and variance parameters. We discuss physically motivated priors in Section~\ref{sect:priors}.

\subsection{Emergent constraints and exchangeable reality} 

The standard procedure in the emergent constraints literature is to assume reality, $y^*$, follows the same regression as the other models. From the statistical view we have given, this implies that $y^*$ is assumed to be exchangeable with all of the climate models given $x^*$. Usually $x^*$ is taken to be the observed predictor (though \citet{coxetal18} integrate out variability in $x^*$ and \citet{bowmanetal18} provide a framework that includes modelling $x^*$ explicitly, as we will later), and then the regression is used to predict $y^*$ and calculate prediction intervals.

Taking the stronger classical version of this assumption first, reality is assumed to be an independent draw from the same distribution that the models were drawn from. This is the strongest possible form of assumption linking models and reality and does not seem defensible, or necessary given that it implies the weaker exchangeability assumption that we shall argue \textit{against} below. 

Rather than assume reality is an independent draw from the distribution of the models, we could assume conditional exchangeability of $y^*$ given $x^*$ with the $y_i$ given $x_i$. This would amount to the view that there are no processes systematically missing from the models, but present in reality, that might cause us to view the behaviour of the real world to be distinguishable from that of the models. \citet{rougieretal13} dismissed this idea out of hand, yet it is the weakest form of the key assumption driving the calculations currently performed for emergent constraints. We propose a general framework to aid our discussion of the issues.

Suppose we believe the physical insight behind the linearity assumption for our emergent constraint so that 
\begin{equation}
  \label{reality}
  y^* \mid x^*, \tbeta^*, \sigma^{*2} \sim \mathrm{N} (\tbeta^{*T} \tx^*, \sigma^{*2}) 
\end{equation}  
was a sensible model, but the regression coefficients and the error standard deviation, $\tbeta^*$ and $\sigma^*$ were uncertain. Suppose, further, that we believe that the relationships across the models are informative for the relationships in reality, but not necessarily the same. A natural way to express this through a statistical model is to state 
\begin{equation}
  \label{reality_beta}
  \tbeta^* \mid \tbeta \sim \mathrm{N} (\tbeta, \Sigma_{\beta^*})
\end{equation}  
and 
\begin{equation}
  \label{reality_sigma}
  \sigma^{*2} \mid \sigma^2 = \sigma^2 + \sigma_R^2, \quad\quad \sigma_R^2 \sim \mathrm{HN} (0, \xi),
\end{equation}  
with $\Sigma_{\beta^*}$ representing ways in which missing or incorrectly parameterised processes across models might change the emergent relationship, and $\xi$ acknowledging structural uncertainty that simply makes us more uncertain about what reality might do even having observed the models. $\mathrm{HN}$ here indicates the Half-Normal distribution which shares the form of the pdf of the traditional Normal distribution, but with support restricted to $\sigma > 0$, \citep{gelman06}. Note that one would be free to change these distributions to incorporate specific physical knowledge where available, but these assumptions are both natural (the reality coefficients are centred on the model coefficients, but uncertain, and the variance for reality is at least as big as the model uncertainty), and sufficient to illustrate a point. 

The current exchangeability between models and reality assumed within the literature is recovered if these extra sources of uncertainty, $\Sigma_{\beta^*}$ and $\xi$ are collapsed to $0$. If even one systematic bias in models, or one missing process known to affect the response can be acknowledged by a researcher or the wider community, then clearly the standard emergent constraints approach is under-reporting uncertainty. We demonstrate this in Section~\ref{sect:example1} using a recently discovered emergent constraint on ECS \citep{coxetal18}.

\subsection{A complete framework for emergent constraints}

We propose to use the extended emergent constraints framework described by the statistical model in equations (\ref{models}) and (\ref{reality}). We discuss general priors for the model parameters, $\pi(\tbeta,\sigma^2)$, in Section~\ref{sect:priors} and in the illustration that follows we shall use the reference prior described above (so our regression for the models will coincide with the classical analysis). We use the model for $\tbeta^*$ given by Equation~(\ref{reality_beta}) and, instead of (\ref{reality_sigma}), we use a folded normal distribution for $\sigma^*$,
\begin{equation}
  \label{reality_sigmaFN}
  \sigma^* \mid \sigma \sim \mathrm{FN} (\sigma, \xi).
\end{equation} 
The Folded Normal is a generalisation of the Half Normal distribution centred on $\sigma$ instead of $0$ with density
\begin{equation*}
  p(\sigma^* \mid \sigma) = \frac{1}{\sqrt{2 \pi \xi}} 
    \left( \exp \left\{ -\frac{1}{2 \xi} (\sigma^* - \sigma)^2 \right\} + 
           \exp \left\{ -\frac{1}{2 \xi} (\sigma^* + \sigma)^2 \right\} \right), 
  \quad \sigma^* \geq 0.
\end{equation*} 
We prefer this to the more natural formulation in (\ref{reality_sigma}) because it doesn't introduce extra parameters, does not bound $\sigma^*$ below by $\sigma$ which may not be appropriate in some circumstances, is strictly positive and tends to the normal distribution when $\sigma$ is relatively larger than $2 \sqrt{\xi}$. Keeping our modelling choices within the Normal family enables researchers to more easily fit their own distributions by thinking about means and standard errors for unknowns. We say more about this in Section~\ref{sect:priors}.

To account for the uncertainty in the observations. Let $x^*$ be the true value of the predictor in reality and $z$ an imperfect observation of it. The simple measurement model
\begin{equation}
  \label{obs}
  z \sim \mathrm{N} (x^*, \sigma_z^2),
\end{equation} 
with given error variance $\sigma_z^2$ accounts for the observation uncertainty. Often this error might be quite large, particularly if the `observation' really comes from reanalysis. To complete the Bayesian model, a prior on the mean $x^*$ should be given. A natural specification is
\begin{equation}
  \label{priormean}
  x^* \sim \mathrm{N} (\mu_x,\sigma_x^2),
\end{equation} 
with the interpretation that, before we see the data, our best guess for real world $x^*$ is $\mu_x \pm 2 \sigma_x$. In situations where $x^*$ must respect physical constraints (e.g. being strictly positive), other distributions can be used, without affecting the generality of the framework, or our methods of inference. Choosing a reference prior $\pi(x^*) \propto 1$ recovers the usual emergent constraints model, and so we use this in our reference calculations throughout.

In any particular problem, we specify the rest of our prior uncertainty through the quantities $\xi$, $\sigma_z^2$, and $\Sigma^*_\beta$ (we shall demonstrate specification of these in our example below and more generally in Section~\ref{sect:priors}). Letting $(Y,X)$ represent the ensemble, we can then use Bayesian software Stan \citep{carpenteretal17} to generate samples from the posterior predictive distribution $p(y^* \mid z,Y,X).$ We give an integral expression for this in Appendix~\ref{appendix:proofs}.

The code we have provided with this paper samples from this distribution and is sufficiently flexible that any of the distributional assumptions we have made (such as the use of Normal and Half Normal distributions) can be easily altered if required. The app we have provided allows users to add their own emergent constraint data and to experiment with the different sources of uncertainty for themselves. What follows is an illustration of these ideas through a re-examination of the \citet{coxetal18} constraint accounting for different levels of uncertainty.

\section{Illustration using a recently found emergent constraint}
\label{sect:example1}
 

We start with the $\Psi$ statistic presented by \citet{coxetal18} as an emergent constraint on climate sensitivity. $\Psi$ is a metric of temperature variability (standard deviation of global temperature divided by the negative root $1$ year lag autocorrelation of temperature), with a given physical justification for why it should have a linear relationship with ECS (though some dispute that justification as part of the discussion to that paper). 

We begin by introducing what we view as sensible uncertainty judgements, adding the uncertainty in layers so that the effects on the constraint can be observed. Throughout, the reference model refers to the standard emergent constraints model computed by sampling from the posterior under the reference prior. Note, throughout, that the reference prior on the regression coefficients ($\pi(\tbeta,\sigma^2) \propto 1/\sigma^2$) with $\pi(x^*)\propto 1$ and with $\xi$, and $\Sigma_{\beta^*}$ in equations (\ref{reality_beta}) and (\ref{reality_sigma}) collapsed to $0$ recovers the usual emergent constraints model.

\begin{figure}[h]
  \centerline{
   \includegraphics[width=0.45\textwidth]{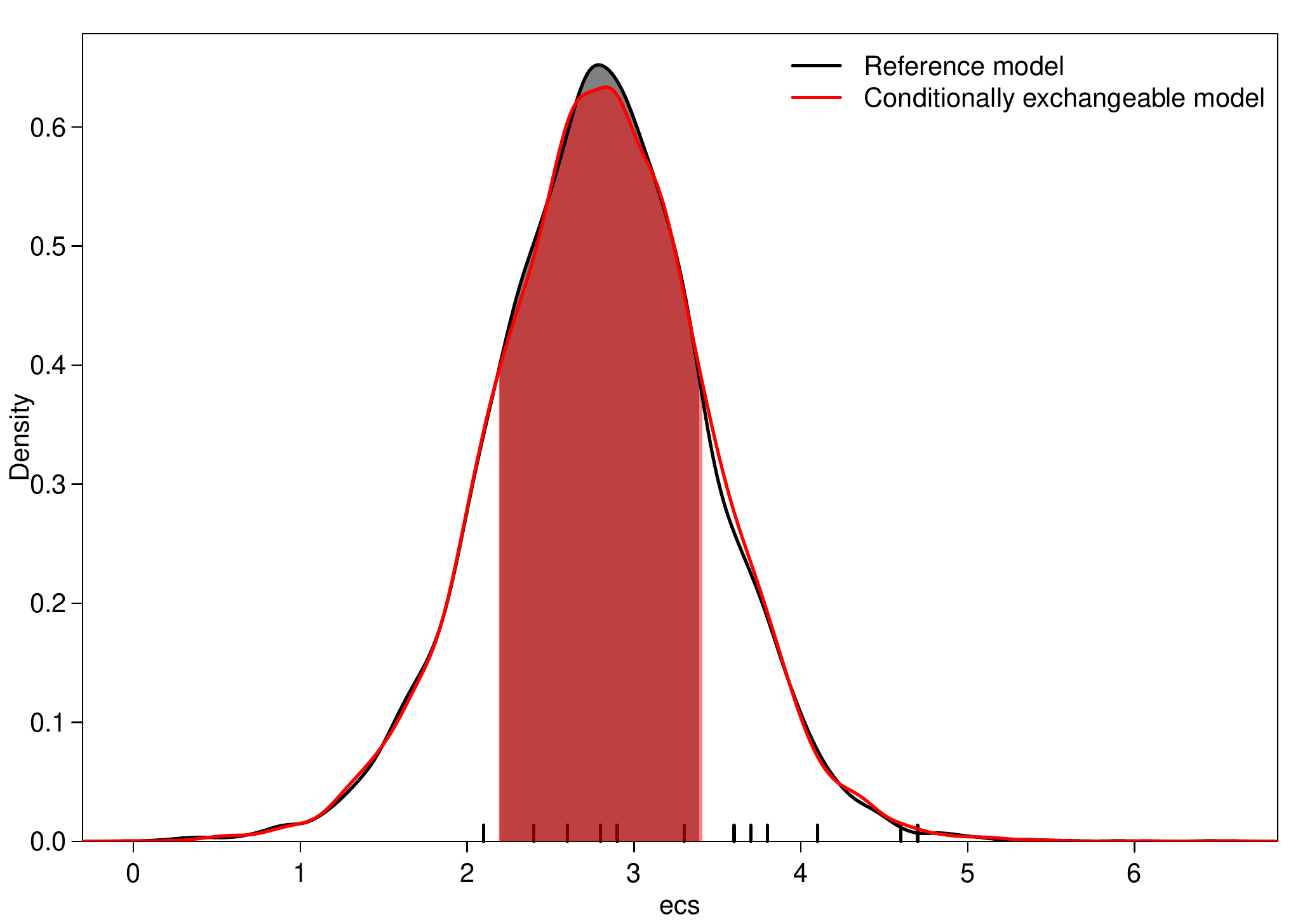}
   \includegraphics[width=0.45\textwidth]{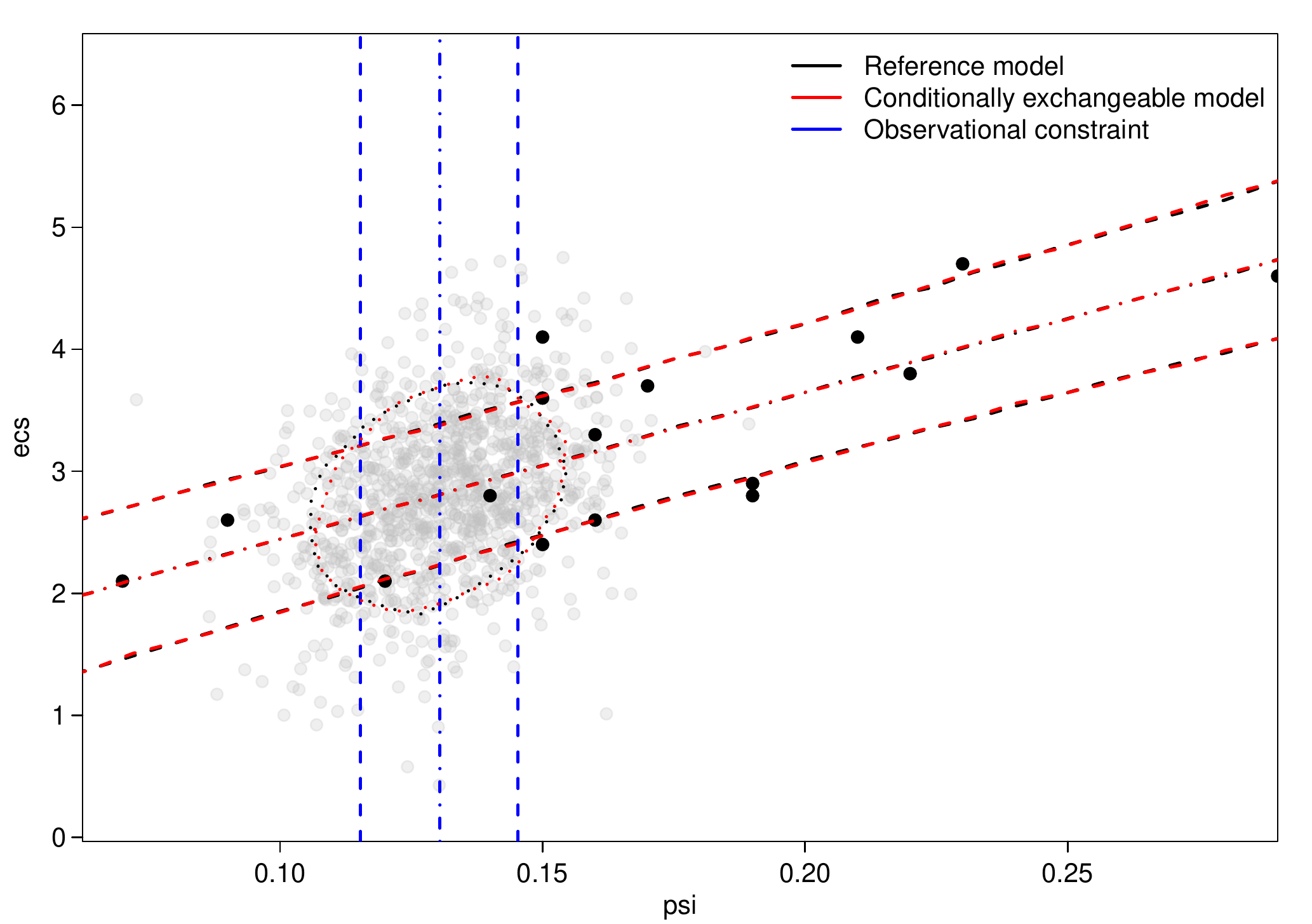}
  }
  \caption{Left: Posterior density for ECS given the models and the observations under the reference prior and with all other uncertainties reduced to $0$ (black) and our model with $x^* \sim \mathrm{N} (0.15, 1)$ (red). The shading represents the $66\,\%$ Bayesian prediction intervals under the 2 models. Right: The Cox constraint vs ECS. Black dots are the CMIP5 models, the grey dots are samples from our posterior distribution for ECS. Blue vertical lines represent the uncertainty on the observation of the Cox constraint and the straight red lines are the median and prediction intervals for the regression relationship for reality. The red and black contours represent the uncertainty on ECS as it depends on the Cox constraint, with black belonging to the reference model and red, our model.}
  \label{figReference}
\end{figure}

We use the HadCRUT4 dataset tabulated in \citet{coxetal18} to give the observations, $z = 0.13$, and their uncertainty $\sigma_z = 0.016$. For our non-reference calculations we set $\mu_x = 0.15$ and $\sigma_x = 1$ based on Figure~2a of \citet{coxetal18} which shows model time series of $\Psi$ (the data is estimated using a moving average approach) across CMIP5, that are all centred between $0.1$ and $0.5$ but with an average of around $0.15$ (by eye). By setting a prior that covers all of the models with much larger uncertainty than an expert may set, we ensure our method is not sensitive to the prior choice (the observation variance is orders of magnitude smaller and so this won't change the posterior very much). Figure~\ref{figReference} shows the posterior distribution of the emergent constraint with these prior choices and reference priors elsewhere. The shading represents the $66\,\%$ Bayesian prediction interval (the probability that ECS is inside the interval is $0.66$, corresponding to the IPCC's ``likely'' range  and chosen to mirror \citet{coxetal18}), with the red curve and shading representing our model with the informed prior on $x^*$ and the black representing the Bayesian reference model that coincides with the usual analysis. The reference model gives the same interval as reported in \citet{coxetal18} $[2.2,3.38]$ (black shading (left plot) and black contour (right plot)). We overlay our model results in red with an interval of $[2.19,3.41]$ and the same median estimate $2.80$.

\subsection{Acknowledging additional uncertainty}

Instead of assuming no uncertainty for $\tbeta^* \mid \tbeta$ and $\sigma^* \mid \sigma$, we look at the effect of adding a ``reasonable'' amount, via specifying non-zero $\Sigma_{\tbeta^*}$ and $\xi$, on the emergent constraint. In Section~\ref{sect:priors} we will offer a principled approach to setting values for these quantities, which will require a number of additional arguments and results. For illustration here, we shall define reasonable in terms of the relationship of these ``reality parameter'' uncertainties to the regression parameter uncertainties that come from the Bayesian model.

Having fit the Bayesian regression, we have our beliefs about the relationship between the models through samples from the posterior $\pi(\tbeta,\sigma \mid Y)$, which can be used to calculate posterior means and standard deviations for the parameters, shown, for the \citet{coxetal18} constraint in Table~\ref{parameters1}. The posterior correlation between $\beta_0$ and $\beta_1$ is $-0.95$.

\begin{table}
  \caption{Posterior means and standard deviations for the model regression parameters.}
  \label{parameters1}
  \begin{center}
    \begin{tabular}{lcc}
      \hline
      Parameter &  Mean & Sd    \\
      \hline
      $\beta_0$ &  1.23 & 0.46  \\
      $\beta_1$ & 12.06 & 2.62  \\
      $\sigma$  &  0.59 & 0.12  \\
      \hline
  \end{tabular}
  \end{center}
\end{table}

\begin{figure}[h]
  \centerline{
    \includegraphics[width=0.45\textwidth]{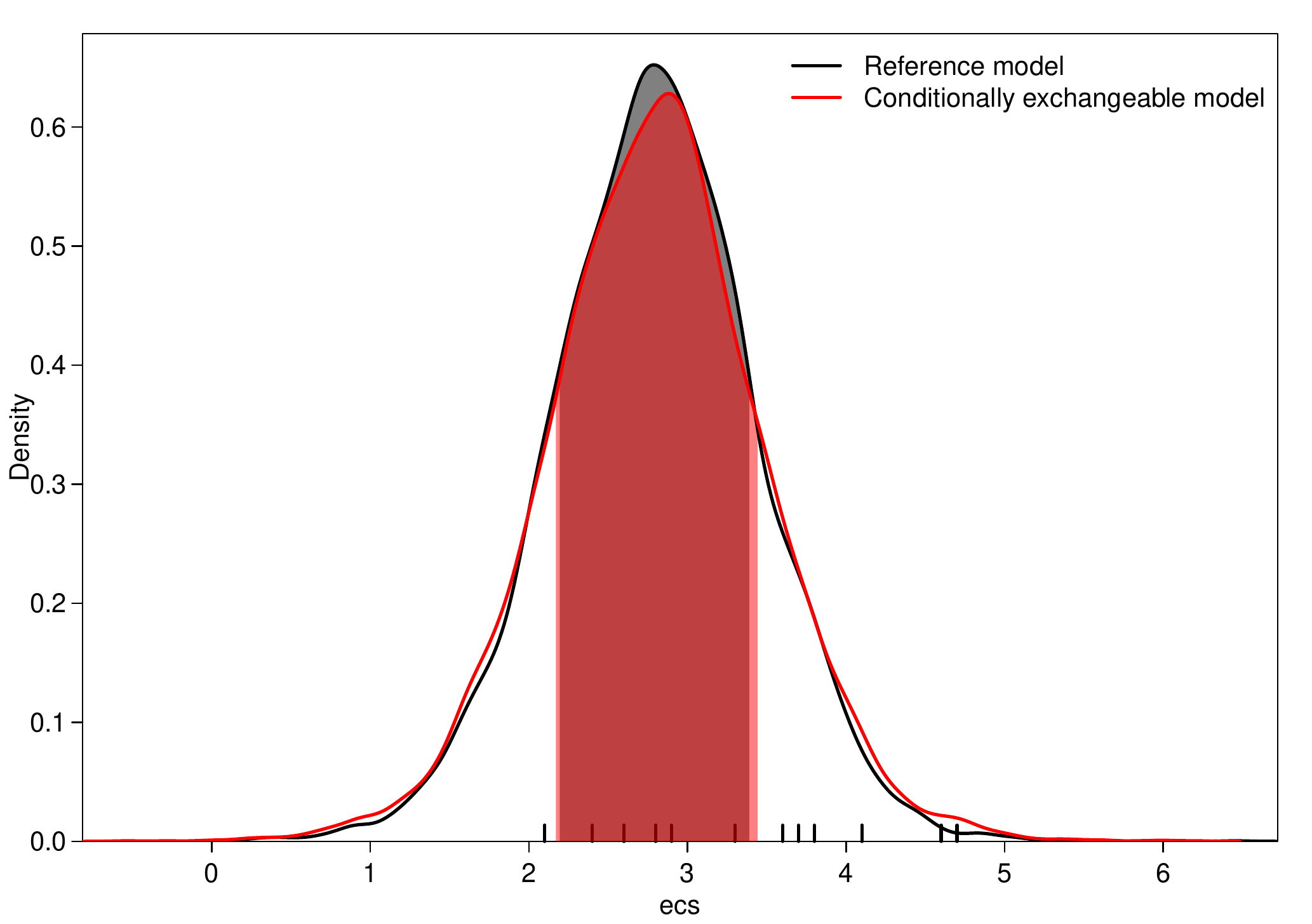}
    \includegraphics[width=0.45\textwidth]{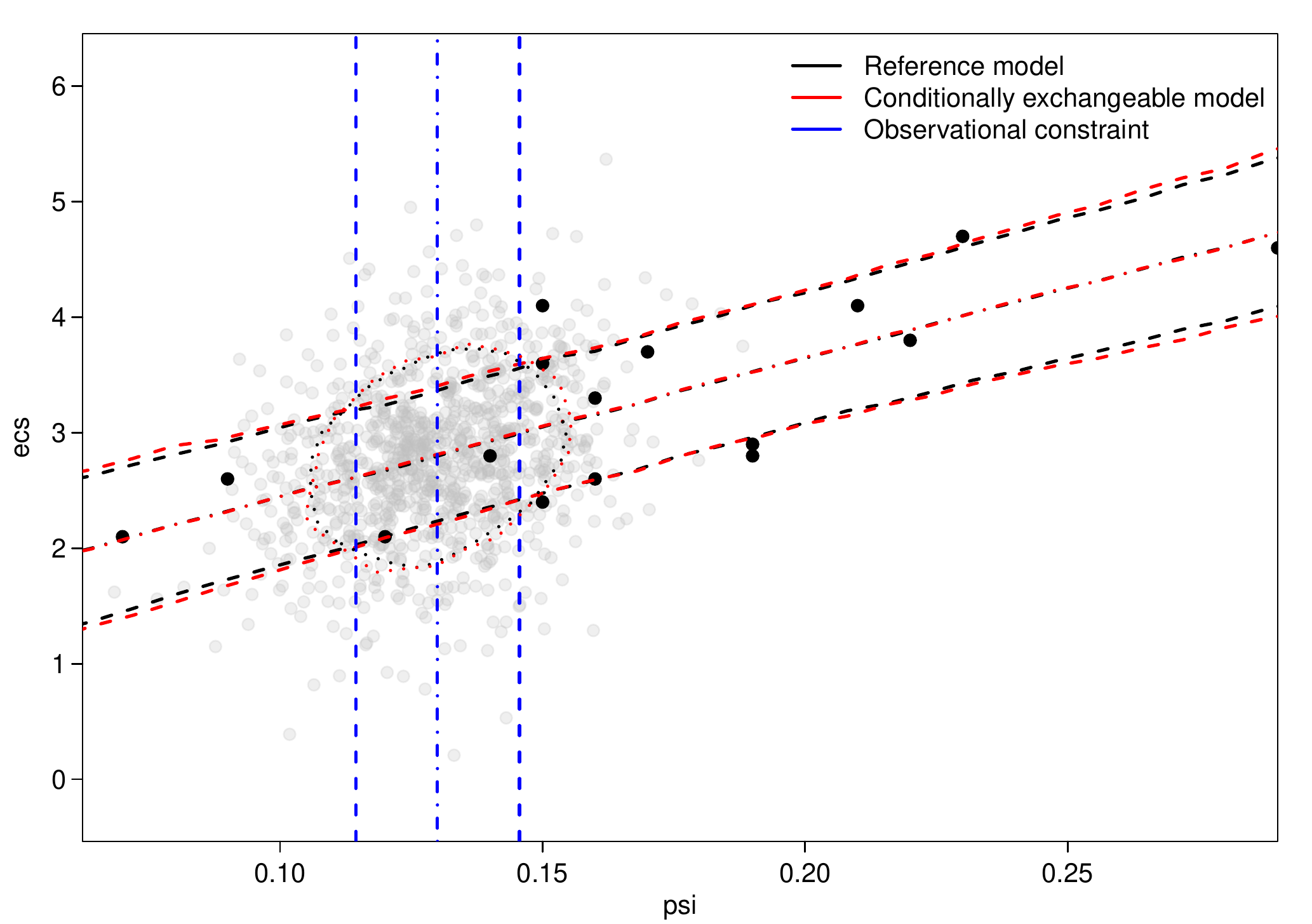}
  }
  \caption{Same format as for Figure~\ref{figReference}, with the posterior uncertainties for the regression parameters adopted for the conditional variances of the reality parameters.}
  \label{uncertainty1}
\end{figure}

We begin with the scenario where, given the values of $\tbeta$ and $\sigma$, we would have the same uncertainty (in terms of standard deviations) for $\tbeta^*$ and $\sigma^*$ as we currently do for $\tbeta$ and $\sigma$, using the numbers in Table~\ref{parameters1} and a correlation of $-0.95$ to construct $\Sigma_{\tbeta^*}$ and $\xi$. This effectively doubles the marginal variance for $\tbeta^*$ and $\sigma^*$. The emergent constraint in this scenario is shown in Figure~\ref{uncertainty1} and has a $66\,\%$ interval $[2.17,3.43].$ We can see from the interval and from the plots that, though we have acknowledged additional uncertainty at a level that may seem reasonable to some, the emergent constraint is hardly changed. Increasing all uncertainties by $10\,\%$ leaves the intervals unchanged (not shown). 

\begin{figure}[h]
\centerline{    
    \includegraphics[width=0.45\textwidth]{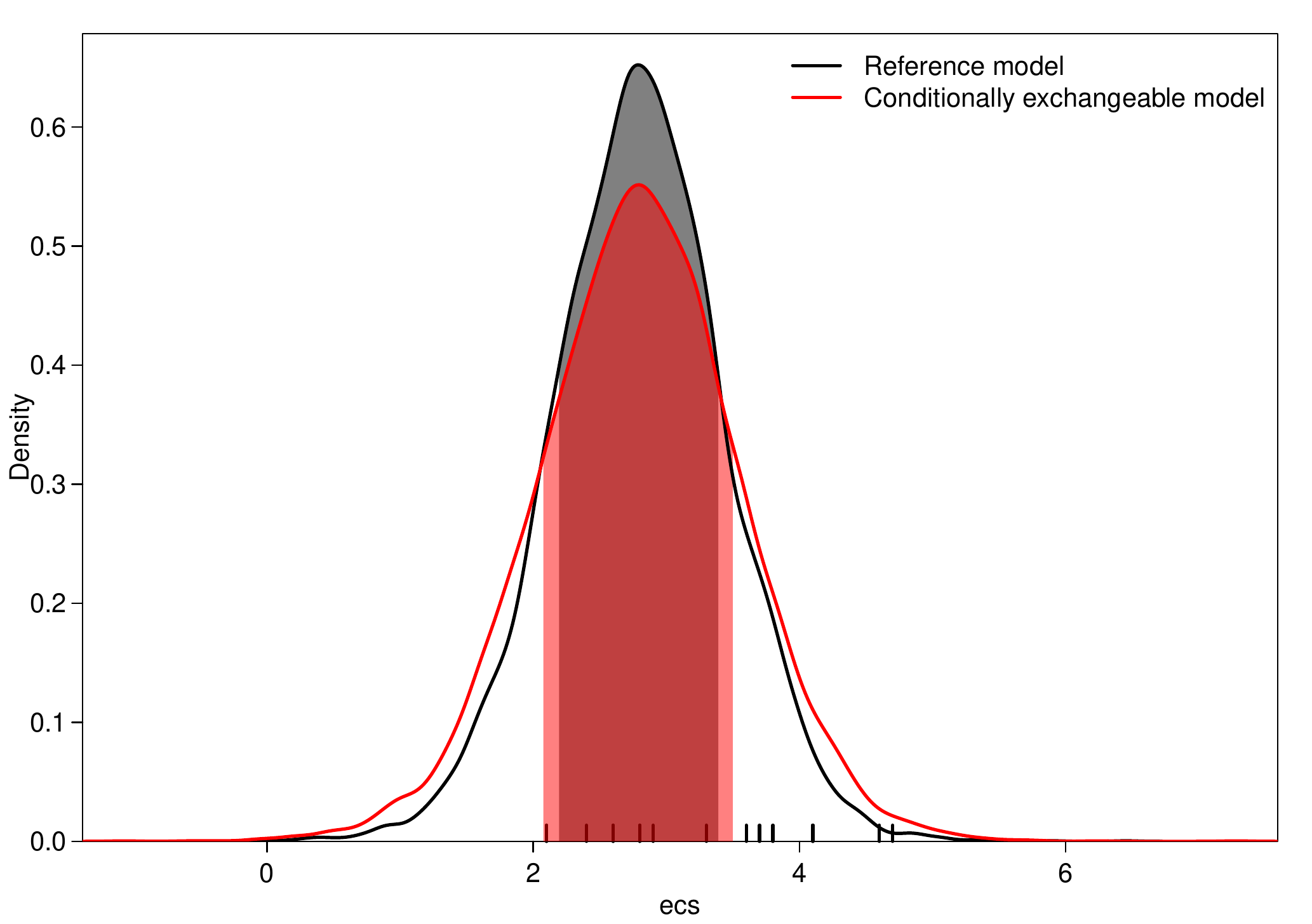}
    \includegraphics[width=0.45\textwidth]{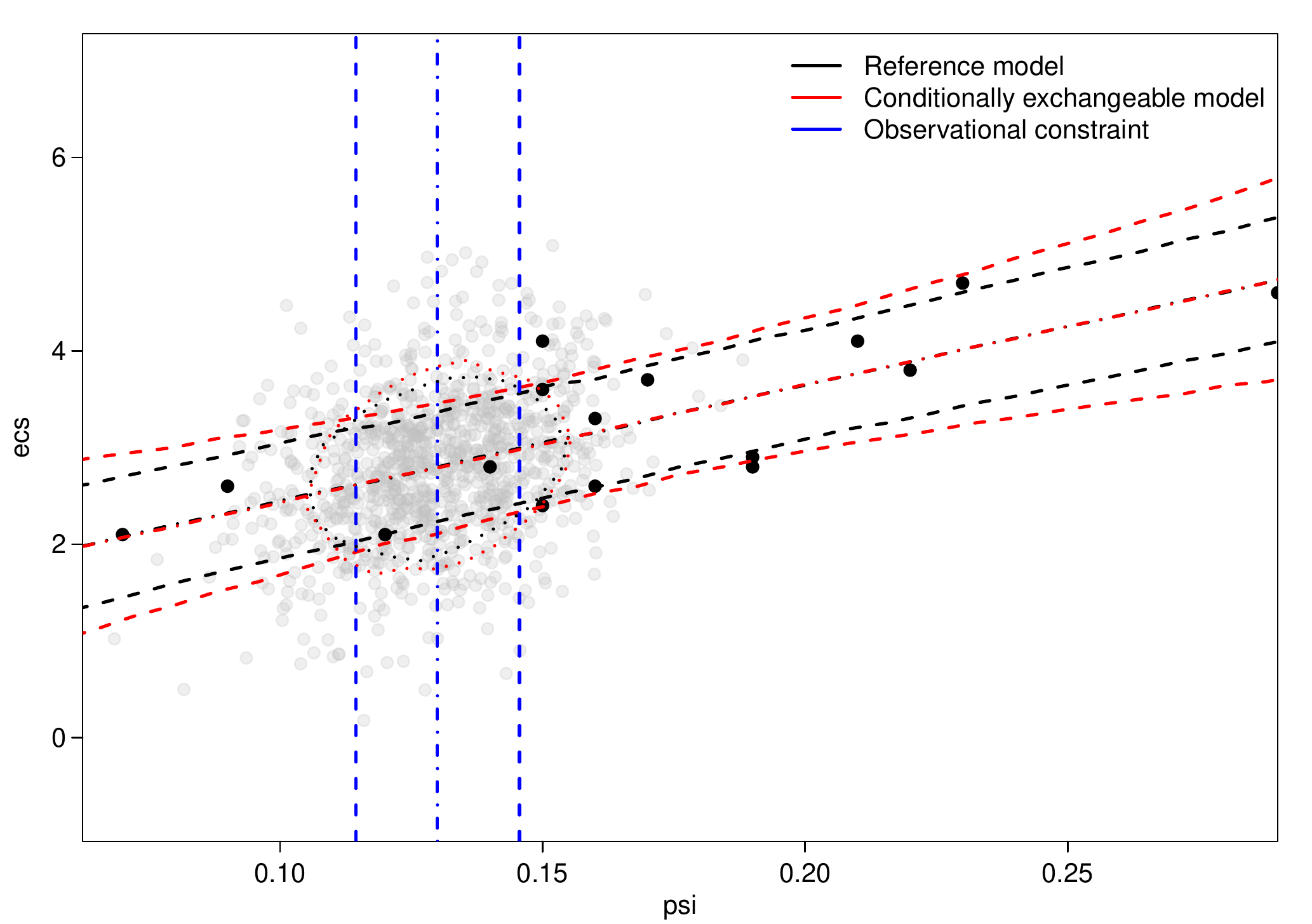}
    }
    \centerline{
    \includegraphics[width=0.45\textwidth]{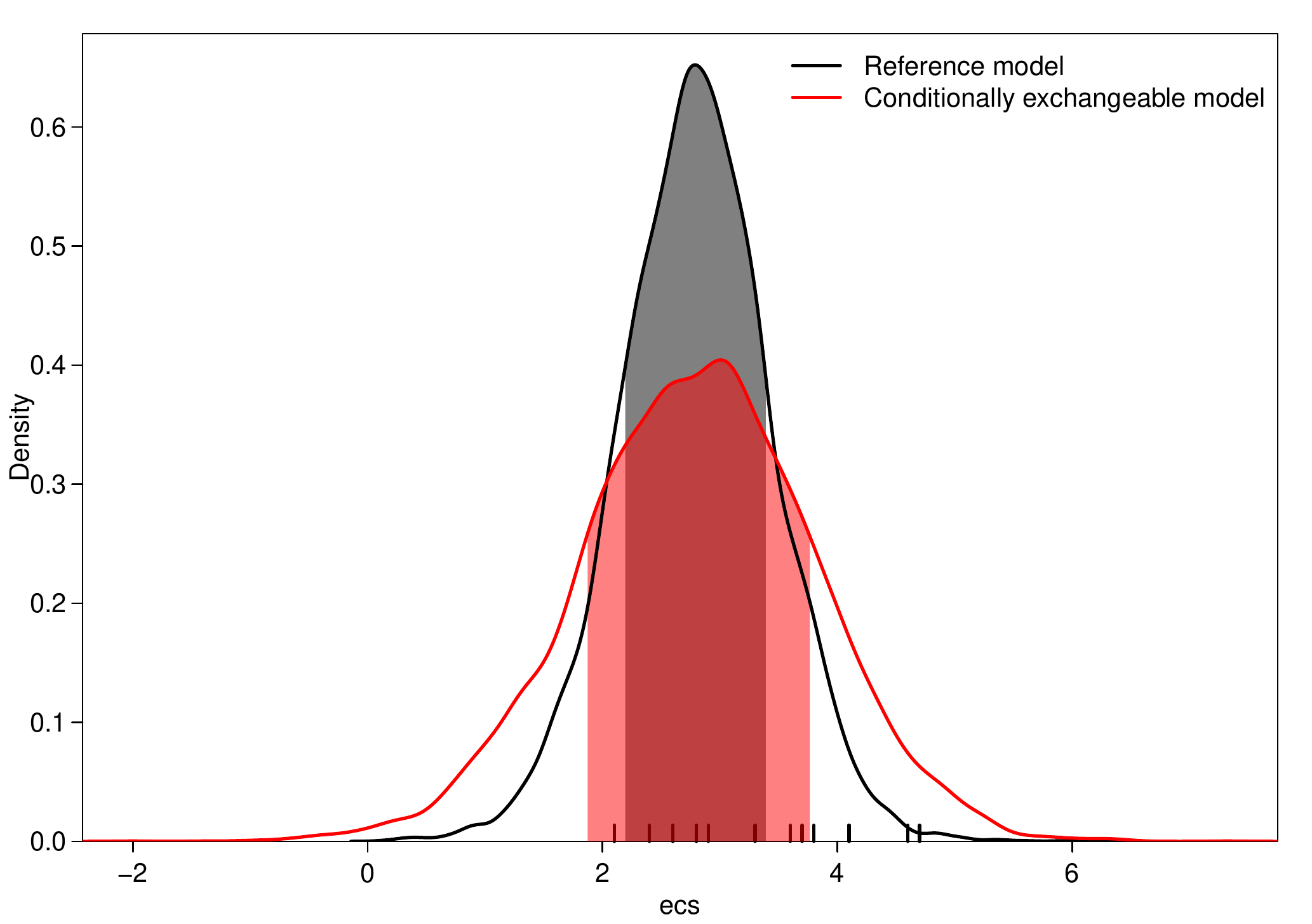}
    \includegraphics[width=0.45\textwidth]{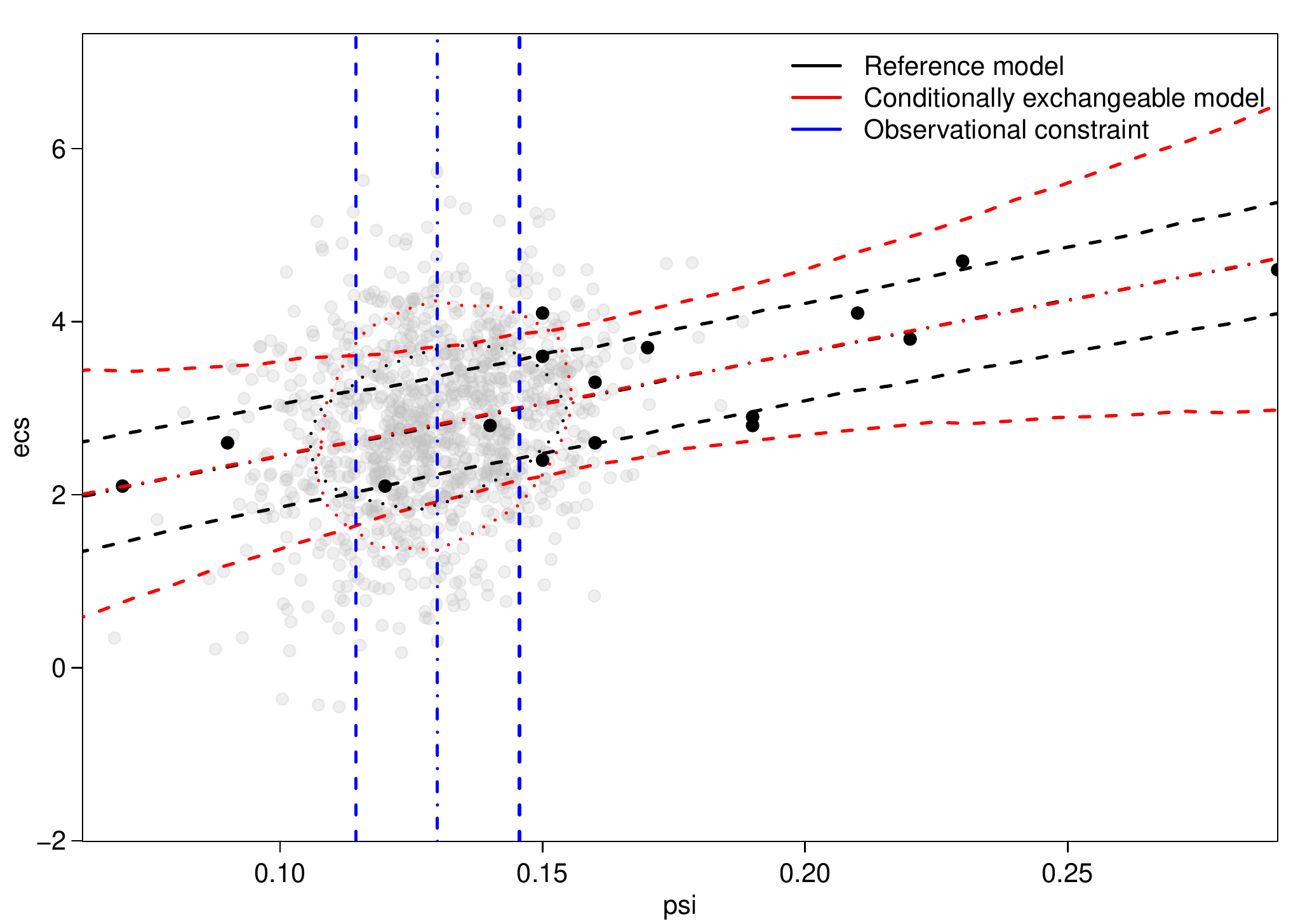}
    }
  \caption{Top row: Emergent constraint plots given a $2.5\,\%$ chance of no constraint. 
           \textit{Bottom row:} Emergent constraint plots given a $16.6\,\%$ chance of no constraint.}
  \label{uncertainty2}
\end{figure}

Note that even with the additional uncertainty specification given above, we are still virtually certain that the emergent constraint exists in reality given the expected value of the models. I.e. our mean for $\tbeta_1^*$ would be $12.06$ and our standard deviation would be $3.71$. For there to be no relationship ($\tbeta_1^*$ crosses $0$) in reality under this model would involve roughly a $4$ standard deviation event, or a probability of $5.76 \times 10^{-4}$! Setting $\sigma_{\tbeta_1}^*$ so that no relationship in reality is a $2$ standard deviation event ($\approx 2.5\,\%$ chance) and a $1$ standard deviation event ($\approx 16.6\,\%$ chance), and setting the standard deviation of $\beta_0^*$ at $1$ and $2$ for these scenarios respectively (based on an argument that says if $\tbeta_1^* = 1$, $\tbeta_0^*$ should be our current best guess for ECS, which we will make more carefully in Section~\ref{sect:priors}), gives $66\,\%$ prediction intervals of $[2.08,3.5]$ and $[1.87,3.74]$ respectively. These constraints are shown in Figure~\ref{uncertainty2} (note we added no additional uncertainty for $\sigma^*$ for these calculations). 

This example shows that not-insignificant additional uncertainty can be acknowledged for an emergent constraint, without dramatically changing the conclusions of the analysis. However, there are clearly sensible levels of additional uncertainty that could matter to an emergent constraint. In any given application, what should the additional uncertainty be? This is a fair question that might often receive the answer ``that depends on the beliefs of the scientist''. Whilst it is hard to argue with this answer and, whilst acknowledging that any firm beliefs of the scientist that can be captured with the parameters above and openly defended should be used, we think there is a place for sensible default settings for these uncertainties that can be used and understood by any practitioner. The risk of not having such defaults is that these real additional uncertainties continue to be swept under the carpet by the community and set to $0$. We present and justify our default choices below.

\section{Confidence-linked default priors for physically motivated constraints}
\label{sect:priors}

The app that accompanies this paper allows the user to work with reference priors throughout and allows all of the quantities that we've introduced by hand to be set manually, giving the user ultimate control and the freedom to express their judgments. For the model regression parameters we go no further than this. In Section~\ref{regression_priors} we describe useful subjective default priors for the regression, but we believe that in many instances ensemble sizes will be sufficient to enable the relatively safe use of the reference prior. For the reality relationships our app offers a third, \textit{guided} specification option, based on the arguments and results from Section~\ref{discrepancy_priors}.

\subsection{Priors for the model relationships}
\label{regression_priors}

Though the reference prior is often deemed the ``objective'' prior choice for regression, it actually imparts far less information than any scientist is capable of. For example, the prior states that all intervals of the same width on the real line are equally likely to contain the true intercept and slope, which is preposterous given even a rudimentary knowledge of the scale of the predictors and responses we might see in the models. Physical knowledge of the response should at least be able to bound the prior support for $\tbeta$ and $\sigma^2$. For example, consider finding an emergent constraint for ECS. We might view it (nearly) impossible that ECS in any model were outside of the range $[0,10]$. So if there were no constraint at all, $\sigma^2$ should be such that the ensemble mean ECS $\pm 3 \sigma$ did not cross both bounds.

A natural choice of prior is 
\begin{equation}
  \label{model_beta}
  \tbeta \sim \mathrm{N} ( \boldsymbol{\mu}, \Sigma_\beta ), 
\end{equation}  
where $\mu_2 = 0$ to ensure that `no relationship' is the most likely outcome a priori and that the sign of the constraint will be dictated by the data. $\mu_1$ can be set to $0$, with $\Sigma_\beta [1,1]$ used to set limits on the prior support for the intercept, or physical arguments such as, ``if the predictor were $0$ what would you expect the response to be'' used to fix these elements of the prior. $\Sigma_\beta [2,2]$ can be used to bound the prior support as discussed above. We would recommend avoiding choosing non-zero off-diagonal entries for $\Sigma_\beta$ \textit{in the prior}, as negative values would indicate a linear relationship was expected and will appear in the posterior when the constraint is estimated if that is the case. 

As argued by \citep{gelman06}, a natural choice of prior for $\sigma$ is a Half-Normal prior, 
\begin{equation}
  \label{model_sigma}
  \sigma \sim \mathrm{HN} ( 0, \sigma_s^2 ),
\end{equation}  
where the Half-Normal distribution shares the form of the pdf of the traditional Normal distribution, but with support restricted to $\sigma > 0$. Though this choice does not lead to analytically tractable Bayesian updating, as with, say, an Inverse-gamma prior, giving a limit to $\sigma_s$ is far easier to do for a user, and modern inference with Stan is extremely fast for problems of this size and type. We apply these ideas to choose a subjective prior for the Cox constraint in the models in Section~\ref{application_priors}.

\subsection{Priors for the real world}
\label{discrepancy_priors}

Equations~(\ref{reality}), (\ref{reality_beta}) and (\ref{reality_sigmaFN}), gave a model for reality $y^*$ as a regression on some predictor $x^*$, with ``reality parameters'' $\tbeta^*$ and $\sigma^*$, that we link to the output of the models. But the interpretation, particularly for $\tbeta^*$ could be problematic. Succinctly, how can there be a regression relationship between $x^*$ and $y^*$ in reality when there is only one reality (one $x^*$ and one $y^*$)? The following construct offers us a way to think about this statistical model.

Suppose, for the generation of models in our ensemble, the values of $\tbeta$ and $\sigma$ could be made known to us (e.g. through many many more models of the current generation being included in the sample). At some future time, an ensemble of the next generation of models will be made available to the community and we can re-examine our emergent constraint, finding $\tbeta'$ and $\sigma'$. We expect the next generation of models to represent physical processes better. Some models will have higher resolution, others will have used the intervening years to develop new parameterisations that overcome known structural biases in their models. If $\tbeta'$ and $\sigma'$, could be made known to us, we would expect them to be different from $\tbeta$ and $\sigma$, as the new physics in the models alters the relationships, even if we may not know if the improved physics would make the slope of the constraint stronger or weaker. We might consider $\tbeta^*$ and $\sigma^*$ to be the model parameters at the limit of the process of improving all of the models and submitting large ensembles. This idea is similar to that introduced as ``reification'' by \citet{goldsteinrougier09}. By considering how different the relationship could be from one generation of models to the next, we may be more easily able to consider the effect of missing processes on the relationship and more comfortably able to conceptualise how/why $\tbeta^*$ might be different from $\tbeta$ (and similar for $\sigma^*$).

We present arguments for sensible default priors for $\tbeta^*$ and $\sigma^*$ that depend on the level of confidence we have in the physical reasoning leading to the existence of the emergent constraint in the models transferring to reality (or the relationship between different classes of models at the conceptual limit of improvement). Our basic argument will be that, for constraints that were effectively data mined using the current ensemble, we should have low confidence in their holding in the real world (or the next generation of models), and for those based on purely physical reasoning we might have a greater degree of confidence. To enable us to talk about our confidence in a constraint given the ensemble and to enable other researchers to make similar arguments or debate the level of confidence that should be present, we require further probabilistic arguments.

Suppose
\begin{equation*}
  \beta^* \mid \beta \sim \mathrm{N} (\beta, \sigma_{\beta^*}^2) \quad\quad \mbox{and} \quad\quad \beta \sim \mathrm{N} (B, \sigma_\beta^2),
\end{equation*} 
then, the marginal distribution for $\beta^*$ is
\begin{equation*}
  \beta^* \sim \mathrm{N}(B, \sigma_\beta^2 + \sigma_{\beta^*}^2).
\end{equation*} 
See Appendix~\ref{appendix:proofs} for a proof of this result. This result is relevant because, given the ensemble, $(X, Y)$, we would expect
\begin{equation*}
  \beta \mid (X,Y) \;\dot{\sim}\; \mathrm{N} (\hat{\beta}, \hat{\sigma}^2). 
\end{equation*} 
This is a well known limiting property of Bayesian analyses for large data and is the basis of the Laplace approximation \citep{gelmanbda}, but is particularly good for this type of quantity. Its veracity can be checked by looking at the posterior samples, a feature available in the software tool that accompanies the paper. Hence, having seen the constraint on the models, we set 
\begin{equation*}
  \beta^* \mid (X,Y) \sim \mathrm{N} (\hat{\beta}, \hat{\sigma}^2 + \sigma_{\beta^*}^2).
\end{equation*} 
A subtle point here is that we are assuming that the model for $\tbeta^*$ in (\ref{reality_beta}) is a prior model conditioned on the ensemble $(X,Y)$ (and it will be similar for $\sigma^*$), rather than a prior we adopt before we see the ensemble. We believe this is the right assumption to use and reflects how emergent constraints research is done in practice. Having found a linear relationship between a predictor and a response in the ensemble (whatever physical arguments led you to look), you must then decide what this tells you about the real world. The posterior mean and standard deviation of the $\beta$s, $\hat{\beta}$ and $\hat{\sigma}$ respectively, are easily computed from the posterior samples, and are provided in the data summary in the accompanying software. It remains to specify $\sigma_{\beta^*}$.

Our ``guided'' expert judgements involve eliciting a researcher's confidence in the constraint holding in the real world (in the sense we made clear above). We use ``confidence'' here in a similar way to the IPCC, and will consider levels ``virtually certain'', ``very likely'', ``likely'', with these words implying the same probability levels as they do in the IPCC ($99\,\%$, $90\,\%$, $66\,\%$). Suppose we have a $100(1 - \alpha)\,\%$ confidence level in our emergent constraint being real. We will interpret this as an interval for $\beta_1^*$ of $[0,T]$ (for a positive constraint), so that the confidence indicates the probability of the constraint crossing $0$ and thus disappearing (we don't need to consider or find $T$). This probability is $P(\beta_1^* < 0) = \alpha/2$. So, for example, suppose you are ``virtually certain'' that your constraint holds in reality, then $\alpha = 0.01$ and $P(\beta_1^* < 0) = 0.005$. Given the Normal marginal distribution for $\beta_1^*$ described above, standard calculations give
\begin{equation*}
  \sigma_{\beta^*}^2 = \frac{\hat{\beta}_1^2}{\Phi(\alpha/2)^{-2}} - \hat{\sigma}^2,  
\end{equation*} 
where $\Phi(\cdot)$ is the cdf of the standard Normal distribution. 

The same mathematics governs the marginal distribution for $\beta_0^*$, however, the same sign changing argument does not work for the intercept. Instead, we consider the effect on the intercept if the slope $\beta_1$ were to change sign. In that case, and as the slope moved through $0$, the intercept should move towards our current expectation for the response. So, for ECS and with a positive constraint, $\beta^*$, as that constraint reduced, the intercept, $\beta_0^*$, should increase and cross our current mean for ECS ($3$, say) at $\beta^* = 0$. Given the confidence in the constraint as above, and a response with current expectation $\mu_{y^*}$, we set $P(\beta_0^* \geq \mu_{y^*}) \geq \alpha/2$, giving
\begin{equation*}
  \sigma_{\beta^*_0}^2 = \frac{(\mu_{y^*} - \hat{\beta}_0)^2}{\Phi(1 - \alpha/2)^{-2}} - \hat{\sigma}_{\beta_0}^2.
\end{equation*} 
We set the correlation between $\beta_0^*$ and $\beta_1^*$ to be equal to that of the models, which in our experience is usually large and negative, reflecting the geometry of fitting straight lines, rather than any particular judgements about the variability.

As discussed in Appendix~\ref{appendix:proofs}, for $\sigma^* \mid \sigma \sim \mathrm{FN} (\sigma, \xi^*)$ and $\sigma \sim \mathrm{FN} (s, \xi)$, the marginal distribution for $\sigma^*$ is 
\begin{equation*}
  \sigma^* \sim \mathrm{FN} (s, \xi + \xi^*).  
\end{equation*} 
This relationship is useful if we find $\sigma \mid (Y,X)$ to be approximately Folded Normal, which we have found to be a reasonable approximation in practice. When it is, we fit $\hat{s}$ and $\hat{\xi}$ using the regression samples numerically. Given a $100(1 - \alpha)\,\%$ confidence level in the constraint (as discussed above), we consider an argument based on the prior uncertainty of the response for fixing $\xi^*.$ If the constraint, $\beta^*_1$, were really $0$, our model for the response would be a mean ($\beta^*_0 = \mu_{y^*}$), as argued previously, with uncertainty around that mean represented by $\sigma^*$. The final judgement our guided elicitation therefore requires is a judgement for how uncertain the response is \textit{currently}, via a standard deviation, $\sigma_{y^*}$. Note that both $\mu_{y^*}$ and $\sigma_{y^*}$, because they pertain to the response (e.g. ECS), could be found via a literature review or even IPCC summaries, as we will use for the Cox constraint.

Having obtained $\sigma_{y^*}$, we set $\xi^*$ using the condition
\begin{equation*}
  P(\sigma^* > \sigma_{y^*}) \geq \alpha/2,  
\end{equation*} 
so that the confidence is linked to whether the constraint actually reduces uncertainty in the response. We set this numerically as there is no analytic expression for the inverse-CDF of the Folded-Normal. In guided elicitation mode, the app that accompanies the paper requires only $\mu_{y^*}$, $\sigma_{y^*}$ and a confidence level in the constraint (any is possible but the defaults use the IPCC levels) to complete the emergent constraints model.

\subsection{Application to the Cox constraint}
\label{application_priors}

Applying the ideas from Section~\ref{regression_priors}, we use the following simple arguments to set $\Sigma_\beta$ and $\sigma_s$. We know from previous IPCC reports that models typically have a climate sensitivity ``around'' $3$ and that an ECS of $10$ or a negative ECS would be hugely surprising (in a CMIP model). Under a naive assumption that each model ECS was a uniform draw from $[0,10]$ with no emergent signal at all, the regression should fit a mean of around $5$ with no slope and the residual standard deviation, $\sigma$, should be around $2.5$ (so that $2$ standard deviations covers the interval). This is a ``worst-case'' type regression where the data are far more spread than anyone familiar with ECS could possibly expect, and no signal at all. We can therefore set $\sigma_s = 2.5$ as a weakly informative prior on $\sigma$. 

$\Psi$ is order $0.1$ and ECS is order $1$. Hence, as $\Psi$ changes, when multiplied by $\beta_1$, we should still expect a change that is order $1$. Hence, if there is a relationship, $\beta_1$ should not be more than order $10$. To be cautious and only weakly informative, we set $\Sigma_\beta[2,2] = 34^2$ so that a $\beta_1$ value of order $100$ is a 3 standard deviation event. Note the expectation is $0$ and so, in the prior, a negative relationship is as likely as a positive one. It is only the magnitude of the possible relationships that we control.

Given changes in ECS that are order $1$ at most, we would expect the intercept, $\beta_0$ to be order $1$ for ECS. To allow for the possibility of strong negative effects, we set a very cautious $\Sigma_\beta [1,1] = 5^2$, so that the event that the absolute value of the intercept is greater than $10$ is a $2$ standard deviation event. Prior predictive checks (available in the app) show that this prior on the models offers a huge range of potential ECS and relationships, whilst being sensible. The posterior in this case is almost identical to the reference analysis, perhaps because the signal is clear and the ensemble is sufficiently large.

For the guided real world uncertainty specification, we interpret the IPCC ``likely'' range for ECS of $[1.5,4.5]$ as implying a central estimate of $\mu_{y^*} = 3$ and a standard deviation of $\sigma_{y^*} = 1.5$. Table~\ref{CoxTable} shows the $66\,\%$, $90\,\%$ and $95\,\%$ prediction intervals under four different confidence levels. What we refer to as ``coin flip'' is $50\,\%$ confidence level, though we use $50.1\,\%$ to avoid numerical issues in our estimation procedure. We say more about this option in the discussion.

\begin{table}
  \caption{Bayesian prediction intervals for ECS using the \protect\citet{coxetal18} emergent constraint with 4 different confidence levels in the physical arguments behind the constraint.}
  \label{CoxTable}
  \begin{center}
    \begin{tabular}{lccc}
      \hline
      Confidence & $66\,\%$ Interval & $90\,\%$ Interval & $95\,\%$ Interval \\
      \hline
      Virtually Certain & $[2.19,3.43]$ & $[1.59,4.04]$ & $[ 1.23,4.36]$  \\
      Very Likely       & $[2.09,3.53]$ & $[1.36,4.28]$ & $[ 0.91,4.70]$  \\
      Likely            & $[1.81,3.79]$ & $[0.81,4.75]$ & $[ 0.20,5.35]$  \\
      Coin flip         & $[1.56,4.05]$ & $[0.32,5.25]$ & $[-0.45,5.97]$  \\
      \hline
    \end{tabular}
  \end{center}
\end{table}

\begin{figure}[h]
  \centerline{
    \includegraphics[width=0.45\textwidth]{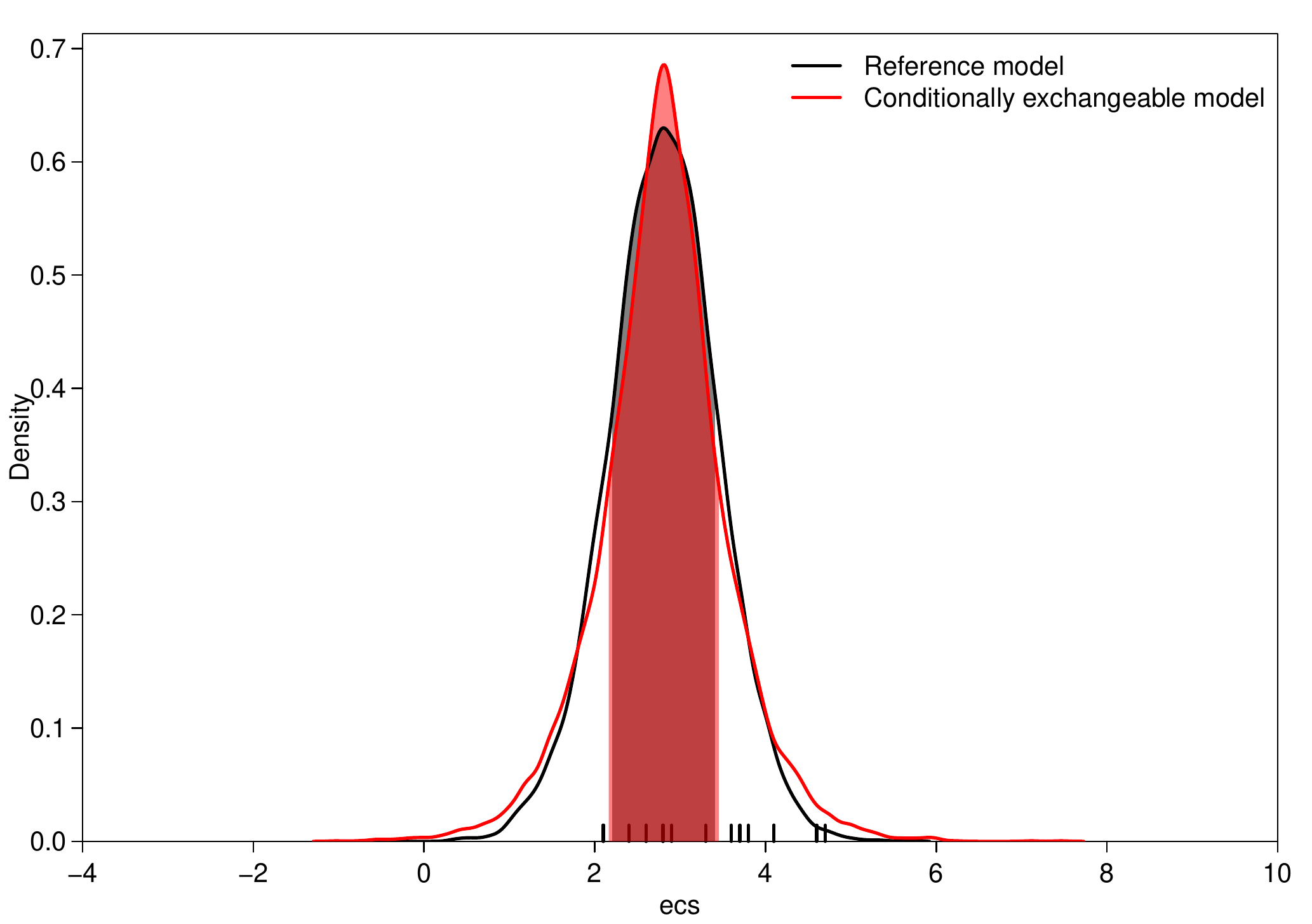}
    \includegraphics[width=0.45\textwidth]{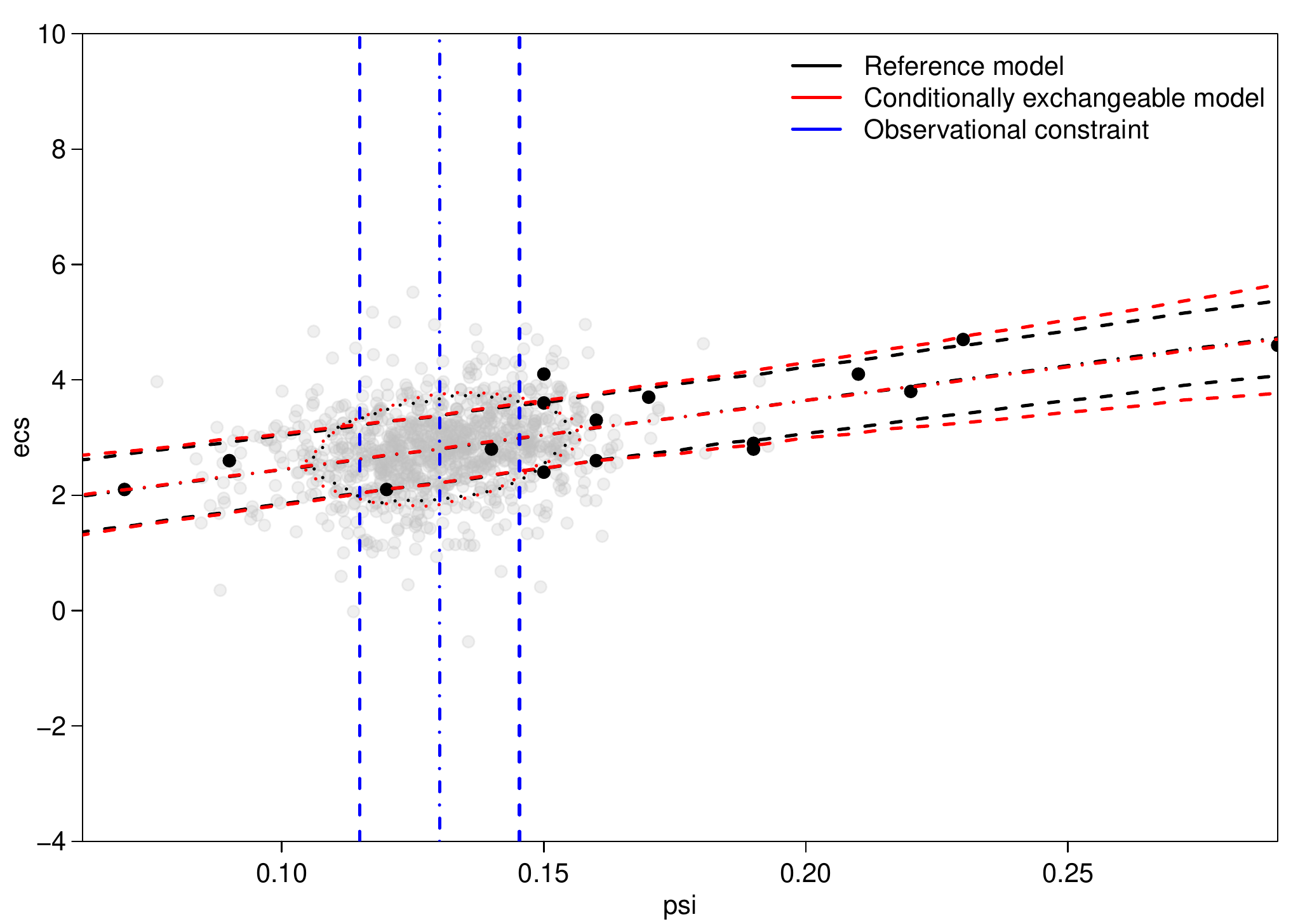}
  }
  \centerline{
    \includegraphics[width=0.45\textwidth]{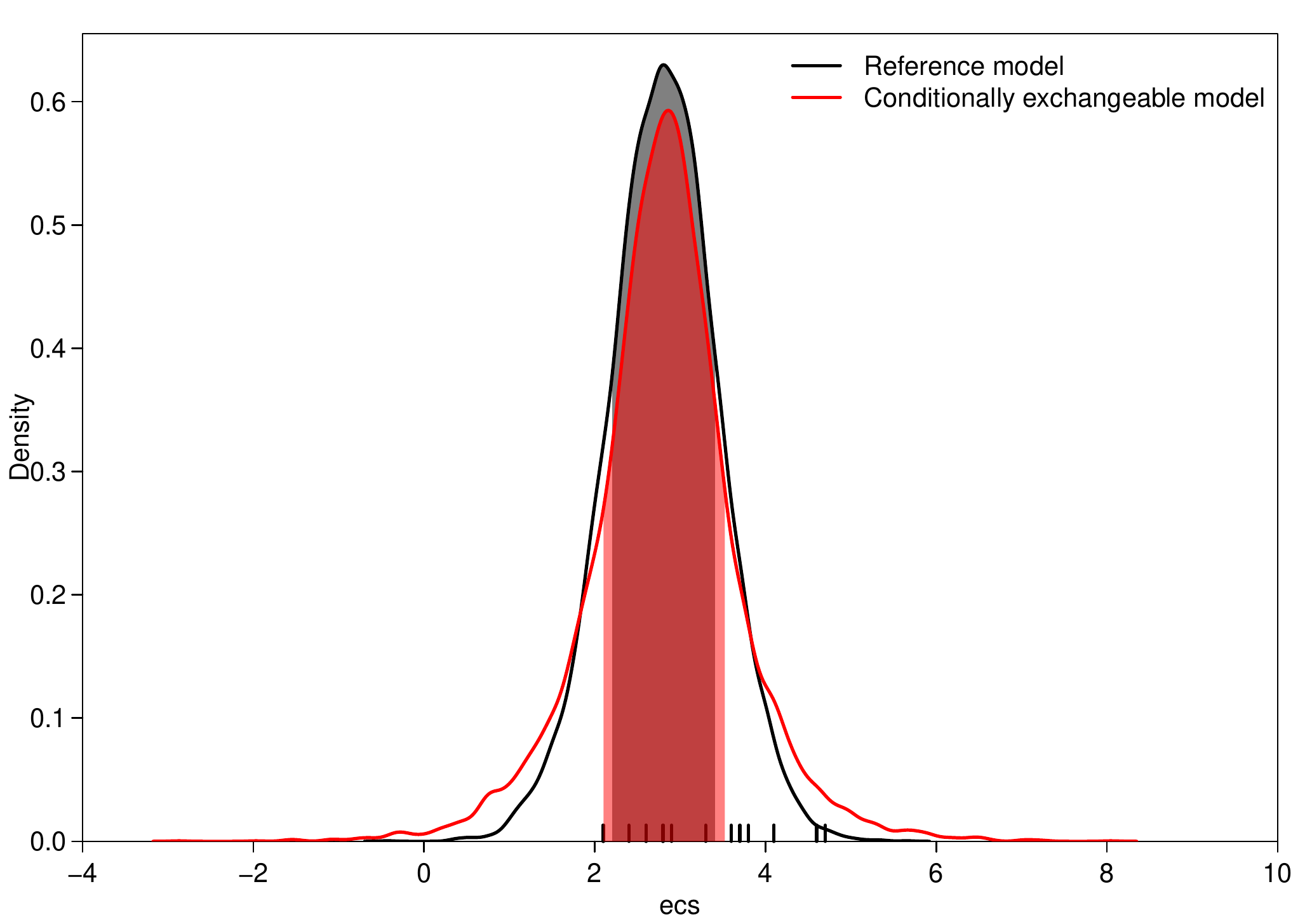}
    \includegraphics[width=0.45\textwidth]{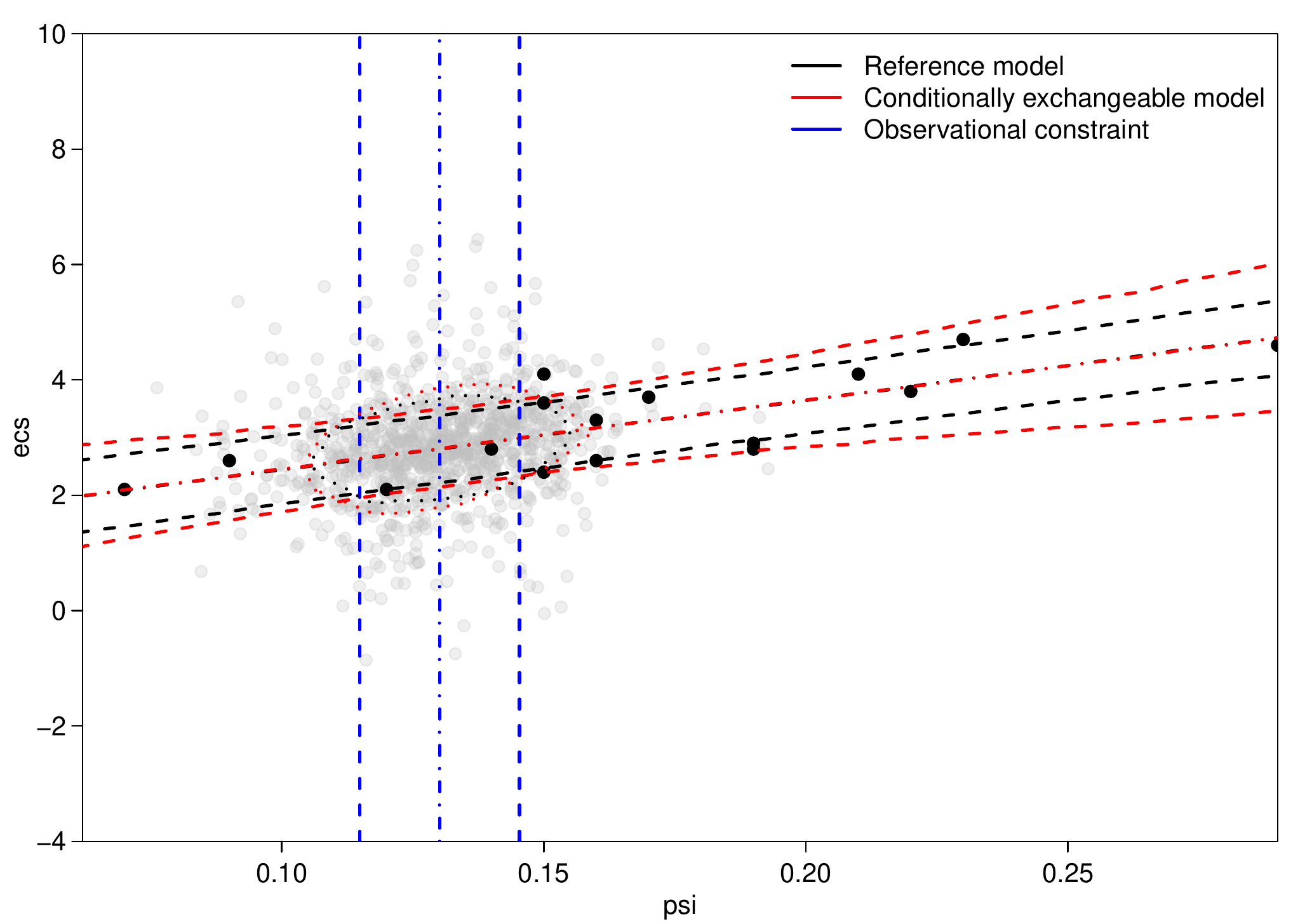}
  }
  \centerline{
    \includegraphics[width=0.45\textwidth]{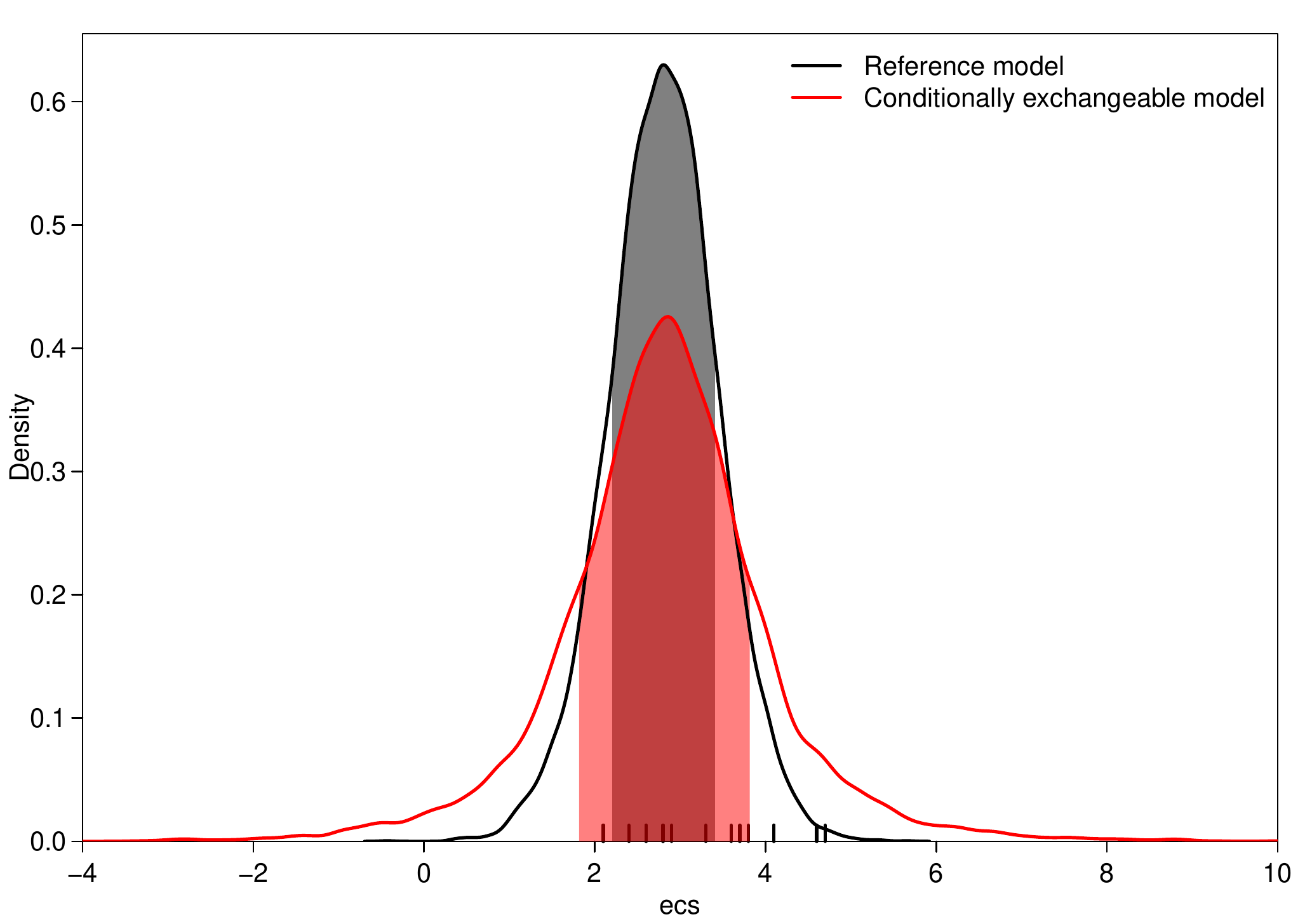}
    \includegraphics[width=0.45\textwidth]{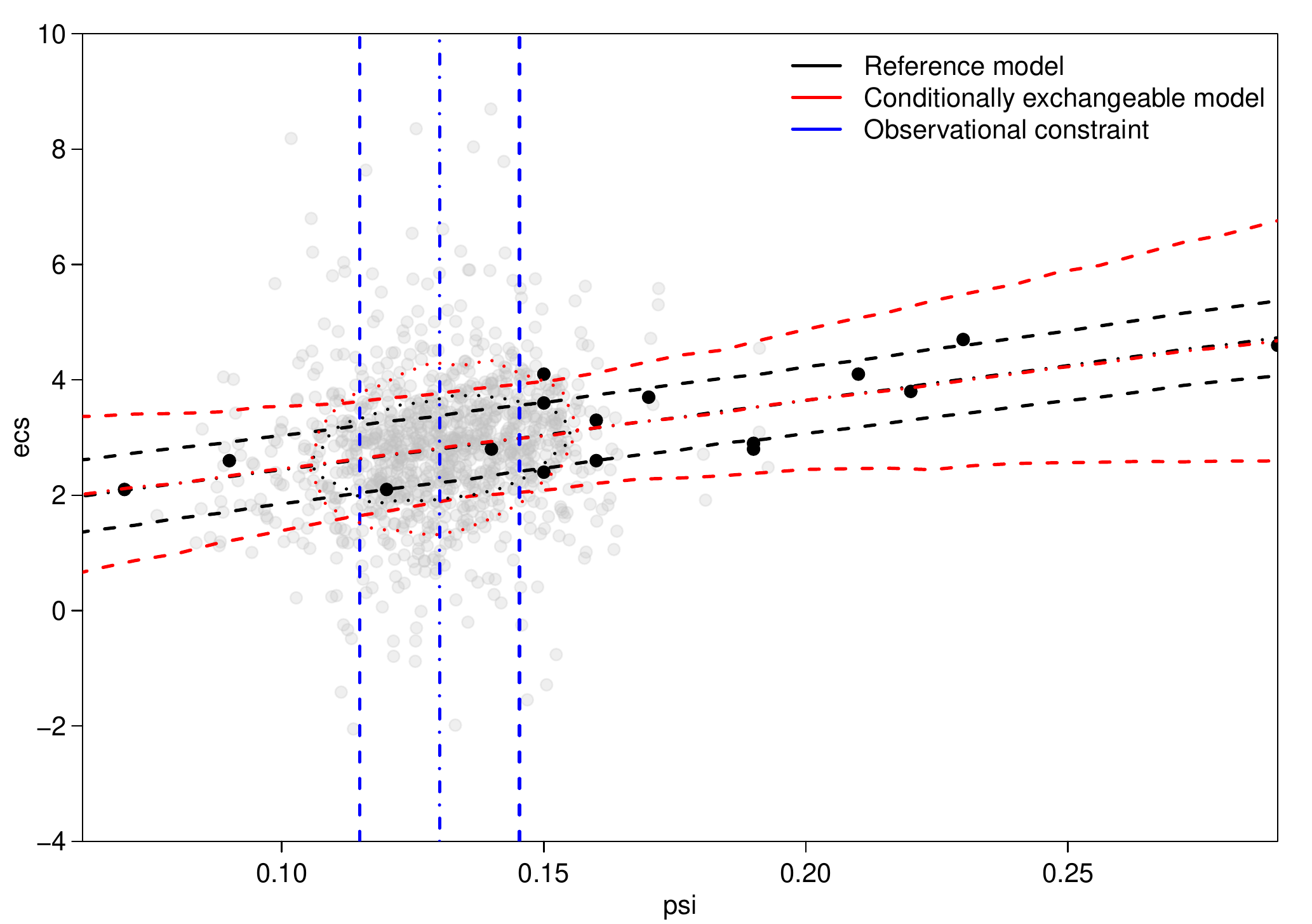}
  }
  \caption{Top row: Emergent constraint plots for ECS given $\Psi$ under a confidence level of \textit{virtually certain} in the existence of the constraint. \textit{Middle row:} As with the top under a \textit{Very likely} confidence level. \textit{Bottom row}, as above with a confidence level of \textit{likely}. The black lines and shading represent the reference model.}
  \label{finalCoxFig}
\end{figure}

The posterior distributions for ECS under the three main levels of confidence are given in Figure~\ref{finalCoxFig}. We see in all cases that acknowledging the additional uncertainty inflates the posterior distribution, but not so much as to remove the constraint. In all cases, having some physical confidence behind the constraint is enough to ensure that something is learned from the analysis. This is even true in the ``Coin flip'' scenario which leads to a note of caution that we expand upon in the discussion: If constraints have been data-mined from an ensemble rather than physically motivated, we do not think this procedure should be used at all. Even fitting the model and specifying some level of confidence requires a strong scientific statement that one must be prepared to back up with physical reasoning. E.g., one consequence of the emergent constraints framework, even our generalised one, is that the central estimate will be determined by the observations and will not be altered by the confidence level.

For the \citet{coxetal18} constraint in particular, we do not offer any judgements as to what the confidence in the constraint should be, as we are not physicists. If the physical reasoning is sound, however, we do insist that the reference model, with all legitimate reality uncertainties ignored, is not appropriate.

\section{Emergent constraints in the literature}
\label{sect:examples}

In this section we apply our extended framework to selected emergent constraints for equilibrium climate sensitivity published within the literature. We only select constraints published with respect to CMIP5 models and we do not include CMIP3 results within the constraints, which may lead our reference intervals to differ from those published. The constraints we choose are: the sum of large and small scale indices for lower tropospheric mixing \citep{sherwoodetal14}, the temporal covariance of low cloud reflection with temperature \citep{brientschneider16}, the double intertropical convergence zone bias \citep{tian16}, and the seasonal variation of marine boundary layer cloud fraction with SST \citep{zhaietal15}. The observations and their standard deviations that we used for each constraint are given in Table~\ref{ConstraintsObs}.

\begin{table}
  \label{ConstraintsObs}
  \caption{Observations and standard deviations used in our analyses of 4 emergent constraints from the literature.}
  \begin{center}
    \begin{tabular}{lcc}
      \hline
      Constraint & Observation ($z$) & Standard deviation ($\sigma_z$) \\
      \hline
      Sherwood et al. (2014)    & $0.825$  & $0.072$  \\
      Brient + Schneider (2016) & $-0.96$  & $0.22$   \\
      Tian (2015)               & $1$      & $0.5$    \\
      Zhai et al. (2015)        & $-1.285$ & $0.565$  \\
      \hline
    \end{tabular}
  \end{center}
\end{table}

\begin{figure}[h]
  \centerline{
    \includegraphics[width=\textwidth]{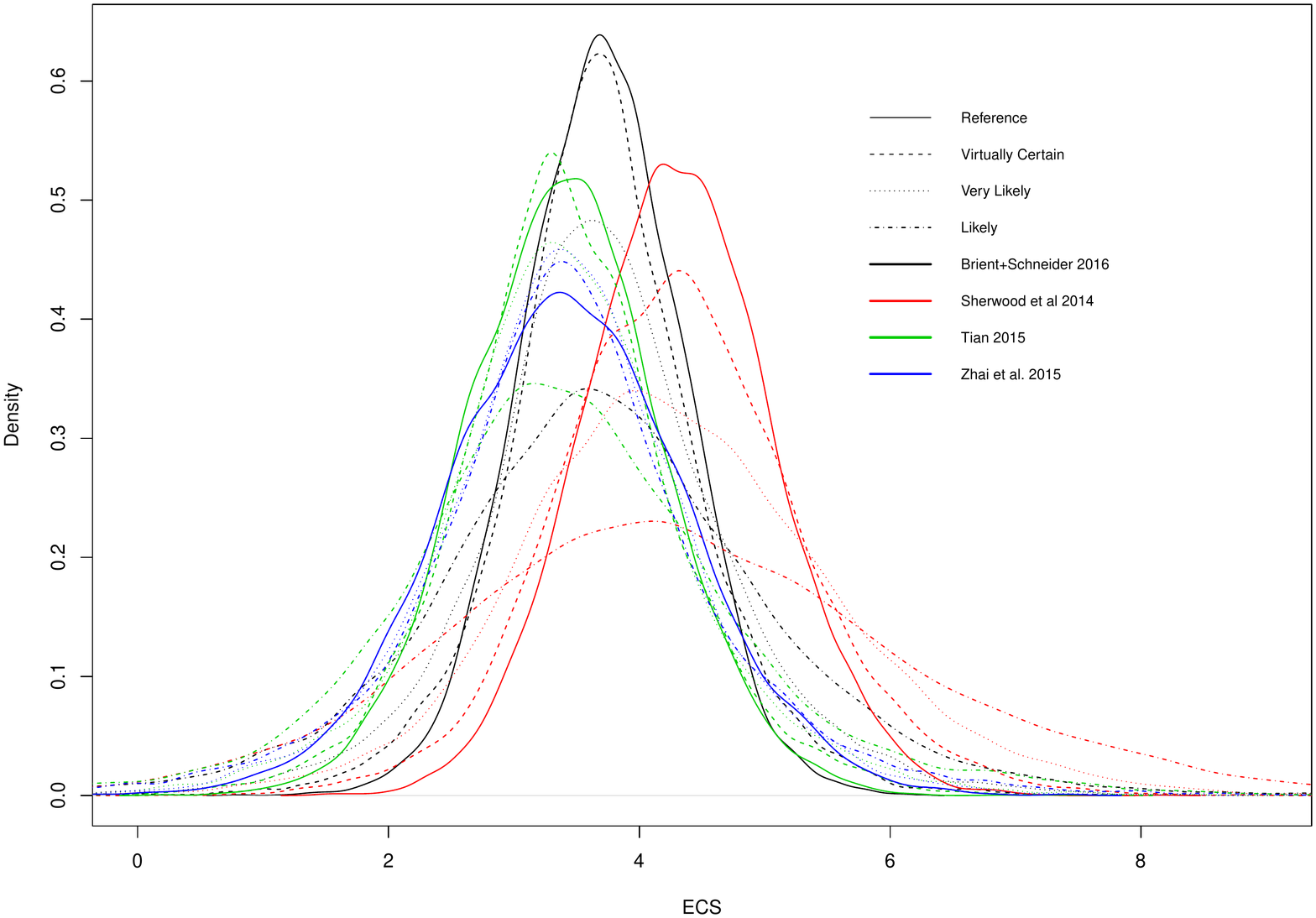}
  }
  \caption{Posterior probability density functions for Equilibrium Climate Sensitivity found for 4 different emergent constraints (colours) and 4 different levels of confidence in the constraint. The solid line in each case is the reference analysis.}
  \label{otherConstraints}
\end{figure}

The results of applying our extended framework for emergent constraints to these data are given as $66\,\%$ prediction intervals in Table~\ref{ConstraintsTable}, and shown as pdfs in Figure~\ref{otherConstraints}, for different levels of confidence in the physical arguments behind the constraints. From the figure we see that in cases where we weaken the confidence in the constraint but where the $66\,\%$ interval remains relatively unchanged, the effect of the additional uncertainty has been to inflate the tails so that our probability of extreme ECS has increased.

\begin{table}
  \label{ConstraintsTable}
  \caption{Bayesian $66\,\%$ prediction intervals for ECS for different published emergent constraints using the reference model and 3 different confidence levels in the physical arguments behind the constraint as per our extended framework.}
  \begin{center}
    \begin{tabular}{lcccc}
      \hline
      Constraint & Reference Interval & Virtually Certain & Very Likely & Likely \\
      \hline
      Sherwood et al. (2014)    & $[3.59,5.02]$ & $[3.42,5.18]$ & $[3.19,5.48]$ & $[2.61,6.04]$  \\
      Brient + Schneider (2016) & $[3.10,4.31]$ & $[3.06,4.37]$ & $[2.90,4.55]$ & $[2.52,4.88]$  \\
      Tian (2015)               & $[2.66,4.14]$ & $[2.65,4.17]$ & $[2.56,4.25]$ & $[2.27,4.54]$  \\
      Zhai et al. (2015)        & $[2.52,4.33]$ & $[2.52,4.33]$ & $[2.53,4.30]$ & $[2.48,4.35]$  \\
      \hline
    \end{tabular}
  \end{center}
\end{table}

We have compared these analyses on alternative emergent constraints on ECS for two reasons. First, to show that the effect of acknowledging reasonable doubt into the existence of each constraint, as discussed via the method of Section~\ref{discrepancy_priors}, is to inflate the prediction intervals, but by a small amount rather than an amount that points to no result. We can say that emergent constraints \textit{have} under-reported uncertainty in the past, but through the given framework, in the future they need not so long as researchers are willing to state their confidence in the underlying physical argument for the linear relationship.

Our second reason is to highlight that published constraints can lead to quite different probability distributions over ECS (e.g. Sherwood predicts a higher climate sensitivity and Cox predicts a much lower climate sensitivity), and to make it clear that these distributions are not compatible in any sense. In each analysis, the authors have made (implicitly) quite different and incompatible conditional exchangeability judgements for ECS given their individual predictors, leading to different models that capture residual variability as Normal with zero-mean. A meta-analysis or review of this literature for ECS that sought to give an idea of the current uncertainty in ECS itself, might stray into somehow combining these intervals or central estimates to give an objective view of the state of the science. This would be particularly troublesome if that combination put more weight on intervals that overlapped. Each interval must be thought of as the scientific judgements of the author, based on his confidence and a transparent set of statistical assumptions, as outlined in Section~\ref{Sect:theory}. A form of meta-analysis might seek to take the individual judgements of a group of scientists and summarise them, but that would not lead to an ``objective'' uncertainty assessment for ECS, but rather an honest survey of the opinions of different scientists asserted with perhaps differing levels of confidence and based on transparent assumptions and beliefs.

\section{Discussion}\label{sect:discussion}

In this paper we sought to unwrap the underpinning statistical assumptions behind the use of emergent constraints to quantify uncertainty for key unknowns in the climate system. We discussed the strong foundational assumptions underpinning the usual classical regression analysis and the interpretation of the real world as a random sample from the distribution of models. We argued that these ideas were too difficult to defend objectively.

We presented the Bayesian view of emergent constraints, and the far weaker and more reasonable a-prior conditional exchangeability judgements that would lead to regression analyses that coincided with the classical analysis under reference priors and showed how, under this framework, standard emergent constraints analyses ignored the key uncertainties present when there are potential structural deficiencies in the current generation of models. We presented a generalised framework for emergent constraints that acknowledged these additional uncertainties, yet collapsed back to the standard model when these uncertainties were set to $0$.

Our modelling looks to adopt the prior judgement that the emergent constraint is informative for reality after having observed the ensemble, to avoid incoherent models for reality beforehand and to acknowledge that these judgements should only be made sparingly. We also believe that this is how scientists think about emergent constraints. As one scientist put it to us by email, ``nobody publishes an emergent constraint that doesn't correlate''. 

We presented a guided prior uncertainty specification that links confidence in the physical reasoning for a linear relationship between the response and the constraint to reasonable additional uncertainties through judgements about the response itself which are either simple to specify or generally available through literature review. We have developed a software tool that allows users to do this for themselves, and have ensured that this tool also allows scientists full freedom to specify any levels of uncertainty on any of the parameters that they wish, if they do not want to follow our guided specification. Our tool is simple to use and we will maintain it for the community through github. 

The arguments in this paper make it very clear that strong scientific judgement is implied when linking models to reality, particularly when claiming that a linear relationship between quantities across models indicates a physical relationship. Data mining for constraints may very well lead to a multiple testing problem. A simple numerical experiment can be used to illustrate the point. Generating $430\,000$ (normal) random numbers and stacking these into a matrix with $43$ rows, generates a pseudo ensemble with $43$ members and with no physical links between the $10\,000$ outputs. Looking at the maximum absolute correlation between outputs across the ensemble will usually return correlations between $0.7$ and $0.85$, well above the threshold for relationships for an emergent constraint. To base the strong beliefs required to take this relationship into the real world (in the way we have made clear), on only the discovery of a large correlation cannot be justified. For that reason, even specifying a low confidence in the constraint through our guided framework would still be inappropriate. 

One criticism of emergent constraints is that they are overly simple, ignoring complex non-linearities or interactions with processes that are not yet well understood or resolved by models. We do not fully agree with this criticism. When the linear relationship can be well established through mathematical and physical arguments, the conditional exchangeability judgements we have explained in this paper, amounting to indifference over labels and, though appreciating that the relationship will not be \textit{exactly} linear, having no strong judgements as to systematic deviations from it, seem plausible in many situations. Whilst the models and reality themselves may well be more complex, that does not invalidate the statistical model which, rather than making strong statements about how reality/the models actually behave, captures our current knowledge and can be defended on those grounds. Of course, more complex forms of regression could be used within the framework we discuss, but the implied beliefs and the way these will be amended for transferring the constraint from models to reality will be far more complex and difficult to defend.

We hope that by making the required statistical assumptions clear and transparent, the validity of any given constraint, new or existing, can be discussed by the community in terms of the physical reasoning, the reasonableness of the exchangeability judgements and the confidence in the current generation of models and linear relationship for a given quantity. By making software available to the community, we hope to help this debate move forward by allowing different researchers to look at the sensitivity of intervals to these judgements and to form their own views.

%

\section*{Acknowledgments}
This work was funded by NERC grant: NE/N018486/1. The authors would like to thank Ben Sanderson for sharing his data on emergent constraints within the literature. We'd like to thank Peter Cox, Mark Williamson and Femke Nijsse for useful discussions about emergent constraints and for sharing their data. The lead author would also like to thank Michel Crucifix for his encouragement to write this paper.



\bibliographystyle{apa}  

\bibliography{emergent}

\appendix

\section{Mathematical details}
\label{appendix:proofs}

\subsection{Posterior predictive sampling}

The posterior predictive distribution for reality given the models and observations is expressed by the following integral
\begin{eqnarray*}
  p(y^* \mid z,Y,X) 
    & = & \int p(y^*,\tbeta^*,\sigma^*,\sigma,\tbeta,x^* \mid z,Y,X) 
               d\tbeta^* d\sigma^* d\sigma d\tbeta dx^* \\
    & = & \int p(y^* \mid \tbeta^*,\sigma^*,x^*)
               p(\tbeta^*,\sigma^*,\sigma, \tbeta, x^* \mid z,Y,X)
               d\tbeta^* d\sigma^* d\sigma d\tbeta dx^* \\
    & = & \int p(y^* \mid \tbeta^*,\sigma^*,x^*) p(\tbeta^* \mid \tbeta) 
               p(\sigma^* \mid \sigma) p(x^* \mid z) 
               p(\tbeta,\sigma \mid Y,X) 
               d\tbeta^* d\sigma^* d\sigma d\tbeta dx^*.
\end{eqnarray*}
Our software samples from each distribution within this factorisation to provide posterior predictive samples.

\subsection{Bayesian updates}

As argued in Section~\ref{sect:priors}, if
\begin{equation*}
  \beta^* \mid \beta \sim \mathrm{N} (\beta, \sigma_{\beta^*}^2) \quad\quad \mbox{and} \quad\quad \beta \sim \mathrm{N} (B, \sigma_\beta^2),
\end{equation*} 
then, the marginal distribution for $\beta^*$ is
\begin{equation*}
  \beta^* \sim \mathrm{N}(B, \sigma_\beta^2 + \sigma_{\beta^*}^2).
\end{equation*} 
To show this, we have
\begin{eqnarray*}
  p(\beta^*) 
    & = & \int_{-\infty}^{\infty} p(\beta^* \mid \beta) p(\beta) d\beta \\
    & = & \int_{-\infty}^{\infty} 
            \frac{1}{\sqrt{2 \pi} \sigma_{\beta^*}} 
            \exp \left\{ -\frac{1}{2 \sigma_{\beta^*}^2} (\beta^* - \beta)^2 \right \} 
            \frac{1}{\sqrt{2 \pi} \sigma_\beta} 
            \exp \left\{ -\frac{1}{2 \sigma_\beta^2} (\beta - B)^2 \right\} d\beta \\
    & = & A\int_{-\infty}^{\infty} \exp \left\{ -\frac{1}{2} 
            \left( \frac{1}{\sigma_{\beta^*}^2} + \frac{1}{\sigma_\beta^2} \right) 
            \left( \beta^2 - 
              2 \left( \frac{\frac{B}{\sigma_\beta^2} + \frac{\beta^*}{\sigma_{\beta^*}}} 
                            {\frac{1}{\sigma_\beta^2} + \frac{1      }{\sigma_{\beta^*}}}
                \right) \beta 
            \right) \right\} d\beta
\end{eqnarray*}
where \[A = \frac{1}{2 \pi \sigma_{\beta^*} \sigma_\beta}
          \exp \left\{ -\frac{1}{2} \left( \frac{{\beta^*}^2}{\sigma_{\beta^*}^2} +   \frac{B^2}{\sigma_\beta^2} \right) \right\}. \]
Completing the square for the integrand, it becomes proportional to a Normal distribution in $\beta$ and so the integral becomes
\begin{equation*}
  \sqrt{2 \pi} 
  \frac{\sigma_{\beta^*} \sigma_\beta}{\sqrt{\sigma_{\beta^*}^2 + \sigma_\beta^2}}
  \exp \left\{ 
    \frac{1}{2} 
    \frac{(B \sigma_{\beta^*}^2 + \beta^{*} \sigma_\beta^2)^2}
         {\sigma_{\beta^*}^2 \sigma_{\beta}^2}
    (\sigma_{\beta^*}^2 + \sigma_\beta^2) 
  \right\}
\end{equation*}
Combining with the constant, collecting the exponential terms and simplifying gives 
\begin{equation*}
  p(\beta^*) = 
    \frac{1}{\sqrt{2 \pi} \sqrt{\sigma_{\beta^*}^2 + \sigma_\beta^2}} 
    \exp \left\{ -\frac{1}{2(\sigma_{\beta^*}^2 + \sigma_\beta^2)} (\beta^* - B)^2 \right\}
\end{equation*}
proving the result. For the Folded Normal result of Section~\ref{discrepancy_priors}, the technique is the same (not shown), though the integral in that case is removed by expressing the integrand as a term proportional to the pdf of a Folded Normal (rather than a Normal).

\end{document}